%% file: main_4.tex
\documentclass[fleqn,10pt]{SelfArx} 

\usepackage[english]{babel} 
\usepackage[toc,page]{appendix}

\definecolor{color1}{RGB}{0,0,90} 
\definecolor{color2}{RGB}{0,20,20} 

\usepackage{hyperref} 
\hypersetup{hidelinks,colorlinks,breaklinks=true,urlcolor=color2,citecolor=color1,linkcolor=color1,bookmarksopen=false,pdftitle={Title},pdfauthor={Author}}

\JournalInfo{Date: February 2 2021 } 
\Archive{ Version 1.0 } 
\PaperTitle{{\LARGE{Group Quantization of Quadratic Hamiltonians in Finance}}} 

\Authors{Santiago Garc\'ia${}^{1}$} 
\affiliation{\textsuperscript{1}\textit{santiago.garcia22@gmail.com}}

\Keywords{Black-Scholes -- Lie Groups -- Quadratic Hamiltonians -- Central Extensions -- Oscillator -- Linear Canonical Transformations -- Mellin Transform -- Laplace Transform} 
\newcommand{\keywordname}{Keywords} 

\date{\today} 

\Abstract{
The  Group Quantization formalism is a scheme  for constructing a functional space that is an irreducible infinite dimensional representation of the  
Lie algebra belonging to a dynamical symmetry group. 
We apply this formalism to the  construction of functional space and operators 
for quadratic potentials- gaussian pricing kernels in finance. We describe the Black-Scholes theory, the Ho-Lee interest rate model and the Euclidean repulsive and attractive oscillators. 
\vskip 1mm
The  symmetry group used in this work has the structure of a principal bundle with base (dynamical) group a semi-direct extension of the Heisenberg-Weyl group by $SL(2, \mathbb{R})$,
and structure group (fiber) $\mathbb{R}^{+}$.
\vskip 1mm
By  using a  $\mathbb{R}^{+}$ central extension,  we obtain the appropriate commutator between the momentum
and  coordinate operators $[\hat{p},\hat{x}]= 1$ from the beginning, rather than the quantum-mechanical $[\hat{p},\hat{x}]= - i \hslash$. The integral transformation  between momentum and coordinate representations  is the bilateral Laplace transform, an integral transform associated to the symmetry group.
 }

\usepackage{xcolor}
\usepackage{graphicx} 
\usepackage{booktabs} 
\usepackage{dsfont} 
\usepackage{amsfonts}
\usepackage{mathtools}
\usepackage{amssymb}
\usepackage{amsmath}
\usepackage[toc,page]{appendix}
\usepackage{sidecap}
\usepackage{float}
\usepackage{mathtools}
\mathtoolsset{showonlyrefs=true}

\DeclarePairedDelimiterX\braket[2]{\langle}{\rangle}{#1 \delimsize\vert #2}

\DeclareMathOperator{\csch}{csch}

\everymath{\displaystyle}
\usepackage{scalerel}[2016/12/29]
\newcommand\Omeg{\scaleobj{0.8}{\omega}}
\newcommand\Sig{\scaleobj{0.8}{\sigma}}
\newcommand\Alp{\scaleobj{0.8}{\alpha}}

\begin{document}
\flushbottom 
\maketitle
\setcounter{tocdepth}{1}
\tableofcontents 
\thispagestyle{empty} 

\newpage

\input{introduction}
\newpage
\input{numeraire_invariance}

\input{group_quantization_formalism}

\input{quadratic_sl2r}
\newpage
\input{black_scholes_nokets}

\input{linear_interest_rates}

\section{Harmonic Oscillator}\label{OSGROUP}
\input{harmonic_oscillator}

%
\section{Repulsive Oscillator}\label{ROSGROUP}
\input{repulsive_oscillator}
\section{Conclusions}
\vskip 1mm

We have presented a methodology, the Group Quantization formalism,  for constructing a  financial theory from symmetry arguments.
The $WSp(2,  \mathbb{R} )$  group has been used to describe the Black-Scholes model, the Ho-Lee model 
and the harmonic and repulsive oscillators.
\vskip 1mm
The choice of  $\mathbb{R}^{+}$  as the structural group in the Heisenberg-Weyl group ensures the appropriate commutator between the momentum
and  coordinate operators $[\hat{p},\hat{x}]= 1$. This choice is compatible with  the financial interpretation of  $\hat{p}$ as a delta. 
\vskip 1mm
In addition, extending translations by  $\mathbb{R}^{+}$ 
formulates the theory directly in real (calculation) time, without the need to
switch to an Euclidean time  in order to derive pricing quantities.
\vskip 1mm
The case of Black-Scholes has been studied in detail. We have constructed the polarized functions and the position, momentum and Hamiltonian operators both in 
coordinate and in momentum spaces.  The role of variance as cohomological invariant has been identified.
\vskip 1mm
First polarizations for the harmonic oscillator have been derived in phase space and in orthogonal coordinates. A  high polarization was necessary to obtain polarized functions 
in coordinate (price)  space.
\vskip 1mm
Group Quantization generates representations of a financial theory in different functional spaces,  allowing for alternative pricing frameworks
such as the  use of Laplace and Mellin transforms.  
In the case of Black-Scholes and Ho-Lee,
the polarized functions have the meaning of prices of financial instruments, while their meaning for the oscillators is object of active research.
\vskip 1mm
As we have shown in this article, among other interesting features, the Group Quantization formalism
provides naturally the functional constraints (polarization algebra or higher polarization) for deriving the pricing equations.
This makes Group Quantization a versatile methodology 
for constructing a  financial theory from symmetry arguments alone, using solely the
principal bundle structure of a centrally extended Lie group.

\newpage
\begin{appendices}
\input{geometric}

\input{black_scholes_prices_in_laplace_space}

\input{black_scholes_lagrangian}

\end{appendices}
\section*{Acknowledgements}

\input{acknowledgements}


\include{references}
\end{document}

%% file: introduction.tex
\section*{Introduction}\label{INTRO}
\vskip 1mm
The  Group Quantization formalism (\cite{aldaya0}, \cite{wolf11}, \cite{garcia0}, \cite{aldaya22})  is a scheme  for constructing a functional space that is an irreducible infinite dimensional representation of the  
Lie algebra belonging to a dynamical symmetry group. 
\vskip 1mm
This formalism utilizes
Cartan geometries in a framework similar to  the Hamiltonian framework, where, in finance,  the {\it{coordinates}} represent prices, rates ...  and the conjugate momenta operators are the corresponding deltas. 
\vskip 1mm
We apply the Group Quantization  formalism to a modified  $WSp(2,  \mathbb{R} )$ group. 
$WSp(2,  \mathbb{R} )$ is the semi direct product of the two-dimensional
real symplectic group {\footnote{The infinite-dimensional representation theory of   $SL(2,  \mathbb{R})$  was used by Bargmann ({\cite{bargmann1}, \cite{bargmann2}) 
for the description of the free non relativistic quantum mechanical particle.}} 
$Sp(2,  \mathbb{R}) \approx SL(2,  \mathbb{R})$
by the Heisenberg-Weyl group.
Our modification consists in that the embedded 2-dimensional translation subgroup in the Heisenberg-Weyl group has
been extended by  $\mathbb{R}^{+}$, rather than $U(1)$, which is the usual extension group in the physics and mathematical literature.
\vskip 1mm
Our interest in  the $WSp(2,  \mathbb{R} )$ group (called {\it{Group of Inhomogeneous
Linear Transformations}} in \cite{wolf2}) is that this group is the symmetry group of the  second-order parabolic differential equations.
Using the  $WSp(2,  \mathbb{R} )$ symmetry, one can find coordinate systems  and operators that map the 
equations of motion and the corresponding solutions (\cite{Miller2}, \cite{wolf2}).
\vskip 1mm
The use in finance  of mathematical methods previously developed in in physics is often based in the formal similarities 
between 
the Black-Scholes equation and the  quantum mechanical Schrodinger equation. These formal similarities have been explored both from the point of view of  Lie algebra invariance  (\cite{Miller}, \cite{lie1},\cite{lie2}, \cite{kozlov} ) and  global  symmetries of the Black-Scholes Hamiltonian operator(  \cite{conformal1}, \cite{conformal2},\cite{conformal3},\cite{conformal4}).
\vskip 1mm
However, 
the mathematical properties  of  a wave equation solution such as the Schrodinger equation
are totally different than the properties exhibited by  a parabolic differential equation.
Moreover, hermiticity  and unitarity do not play a prominent role in finance:
unlike the case of quantum mechanics, in finance there is no probabilistic interpretation of solutions of the pricing equation, and time evolution is irreversible 
and non-unitary.
\vskip 1mm
The Group Quantization formalism makes the most of the principal fiber bundle structure linked to the central extension of a Lie group. Although this formalism shares some 
features with the {\it{Geometric Quantization}} scheme (\cite{geom1}, \cite{geom2}, \cite{geom3}), contrary to Geometric Quantization,
the   Group Quantization formalism does not require the previous existence of a   Poisson  algebra. In both formalisms, the word {\it{quantization}} signifies {\it{irreducible representation}}.
\vskip 1mm
As an example of the applicability of this formalism to finance, we will obtain the Black-Scholes theory, 
the Ho-Lee model and the Euclidean attractive and repulsive oscillators.  
\vskip 1mm
\subsection{Outline}
\vskip 1mm
Section \ref{NUM} shows that a Galilean transformation on the space of Black-Scholes  solutions constitutes a {\it{numeraire change}}. Strict Galilean invariance plays the  role  of
phase invariance in quantum mechanics.
\vskip 1mm
 We describe the main features of Group Quantization in section \ref{qgroupform}. Particularly important are the definition of the connection form, the polarization algebra and the concept of higher polarization. Group Quantization considers the action of the group on itself, as opposed to the group acting on an external manifold.  This guarantees the existence of two sets of commuting  generators, the right invariant fields and the left invariant fields. Left invariant fields
provide naturally a set of {\it{polarization}} constraints that result in pricing equations, while  the right invariant fields provide operators compatible with these constraints.
 \vskip 1mm
 Section \ref{SL2RGROUP} gives a brief survey of the  $WSp(2,  \mathbb{R})$ group, the $SL(2, \mathbb{R})$  group and the Linear Canonical transformations. By  using a 
 $\mathbb{R}^{+}$ central extension to create the embedded Heisenberg-Weyl group, Group Quantization provides  the appropriate commutator between the momentum
and  coordinate operators $[\hat{p},\hat{x}]= 1$, compatible with  $\hat{p}$ representing a delta, rather than the quantum-mechanical $[\hat{p},\hat{x}]= - i \hslash$,  without resorting to analogies with any quantum theory or rotating to   a fictitious {\it{Euclidean time}}.  
 \vskip 1mm
 Section \ref{BSQUANT} applies the Group Quantization formalism to a parabolic $SL(2, \mathbb{R})$ subgroup in order to construct the Black-Scholes theory. We obtain polarization constraints, operators and pricing Kernel both in the momentum space and in coordinate space. This Heisenberg-Weyl commutator makes the bilateral Laplace transform the integral transform mapping  momentum and coordinate spaces. We give examples of pricing in momentum space in appendix  \ref{BRAKET}.
 \vskip 1mm
 Sections  \ref{IRQUANT} to  \ref{ROSGROUP} discuss the application of  Group Quantization  to theories with {\it{deformed}} Black-Scholes equations, i.e., Black-Scholes with a (at most quadratic) potential.
 \vskip 1mm
The Group Quantization of the linear potential, that in finance represents the Hee-Lo interest rate theory, is discussed in section \ref{IRQUANT}. We obtain the polarized functions and pricing equations both in momentum space and in coordinate space. We compare some of our results with the similarity methods developed in \cite{Miller2} and \cite{wolf2}.
 \vskip 1mm
The harmonic and repulsive oscillators are studied in sections \ref{OSGROUP} and \ref{ROSGROUP}.  The Heisenberg-Weyl commutator $[\hat{p}, \hat{q}] = 1$ implies that the harmonic 
oscillator is generated by a hyperbolic  $SL(2, \mathbb{R})$ subgroup,  a change of scale operator, not a rotation. We discuss in detail the polarization constraints in phase space and in a {\it{real}} analog of the Fock space. In spite of 
our non-standard $[\hat{p}, \hat{q}] = 1$, we recover the usual ladder operators when we construct higher order polarized functions. The Mehler kernel is obtained first by building the Kernel  from
the (Hermite) polarized functions, and later by applying the Linear Canonical Transformation technique described in \cite{wolf2}.
 \vskip 1mm
 From the standpoint of the Group Quantization formalism, the  repulsive oscillator can be obtained from the  harmonic and oscillator by making the
 frequency parameter pure imaginary. Both oscillators, which as an interest rate theory 
represent quadratic interest rates, have also been recently proposed as  models for stock returns ( \cite{hos1}, \cite{repul1}).
 \vskip 1mm
 Appendix \ref{DEF} contains some general definitions in order to make this paper more self-contained and  to establish notation.  
 Appendix \ref{CLASSA} shows examples of the Lagrangian formalism and the relationship
 of the connection form in the Group Quantization formalism and 
 the Poincar{\'e}-Cartan form in classical mechanics.
\vskip 1mm
We have favored clarity over mathematical rigor, and important topics such as group cohomology are just glossed over.
For simplicity, we use coordinates as much as possible instead of a more compact notation. 
We refer the reader to the publications in the bibliography for
a detailed and rigorous treatment of the mathematical concepts  relevant to this article.

%% file: numeraire_invariance.tex
%
\section{Galilean Transformations as Numeraire Change }\label{NUM}
As a motivation for this work, we show that the action of the Galilei group on the space of Black-Scholes  solutions constitutes a {\it{numeraire change}}
\vskip 1mm
The Black-Scholes equation for a stock that pays no dividends is
\begin{equation}
\frac{\partial V}{\partial t}  = - \frac{1}{2} \sigma^2 S^2 \frac{\partial ^2 V}{\partial S ^2}  - r S   \frac{\partial  V}{\partial S} +  r \, V  
\label{bs1m1}
\end{equation}
where
$\sigma$ is the stock volatility, $r$ is the risk free rate, $S$ is  the stock price, $t$ is time, and  $V =   V(S, t)$ is the price of a financial instrument. For simplicity, let's assume that $r$ and $\sigma$ are constant.
Under the change of variables
\begin{equation}
S \rightarrow S^\prime = S e^{v^\prime t^\prime} \qquad t \rightarrow t^\prime 
\label{bs1m2}
\end{equation}
where $v^\prime$ is  constant, the Black-Scholes equation becomes
\begin{equation}
\frac{\partial V^{\prime}}{\partial t^\prime}  = - \frac{1}{2} \sigma^2 {S^\prime}^2   \frac{\partial ^2 V^{\prime}}{\partial {S^\prime}^2}  - 
(r  + v^\prime ) S^{\prime}   \frac{\partial  V^{\prime} }{\partial S^{\prime}} +  r\,V^{\prime}
\label{bs1m3}
\end{equation}
where $ V^\prime = V(S^\prime, t^\prime)$. 
\vskip 1mm
The transformations \eqref{bs1m2} can be expressed in log-stock coordinates as
\begin{equation}
\label{bs1m2b}
x^{\prime }  =  x + v^\prime, \, t \quad v^{\prime } =  v \quad t^{\prime } =  t    
\end{equation}
were $x \equiv \ln(S)$.  We recognize the familiar Galilean transformations, with the {\it{position}} $x$ representing the logarithm of the stock price, and the {\it{velocity}} $v$ the stock's growth rate. 
Given that the action of the Galileo group is linear in the log-stock cooordinates, it is convenient to express  the Black-Scholes equation \eqref{bs1m1}  in terms of the logarithm of the stock price
\begin{equation}
\frac{\partial V}{\partial t}  = - \frac{1}{2} \sigma^2   \frac{\partial ^2 V}{\partial x^2}  - \mu   \frac{\partial  V }{\partial x} +  r\, V
\label{bs1m1b}
\end{equation}
with $\mu = r - \frac{1}{2} \sigma^2$.
Equation \eqref{bs1m3} becomes
\begin{equation}
\frac{\partial V^{\prime}}{\partial t^\prime}  = - \frac{1}{2} \sigma^2   \frac{\partial ^2 V^{\prime}}{\partial {x^\prime}^2}  - 
(\mu + v^\prime )   \frac{\partial  V^{\prime} }{\partial x^{\prime}} +  r \, V^{\prime}
\label{bs1m2}
\end{equation}
where  $ V^{\prime} =   V(x^\prime, t^\prime)$
\vskip 1mm
The Black-Scholes equation is only covariant  under the action of the Galilei group.
Strict  Galilean invariance  can be restored by multiplying  the solution $V^\prime$  by an exponential factor
\begin{align}
\bar{V}  (x^\prime, t^\prime) & =  e^{\epsilon(x^\prime,  t^\prime )} \,  V^\prime (x^\prime, t^\prime) \nonumber \\
& \epsilon(x^\prime, t^\prime )  =  \frac{v^\prime } { \sigma^2}  \left( \frac{1}{2}\, v^\prime \, t^\prime   - \left(  x^\prime - \mu   t^{\prime}  \right) \right)
\label{sch3d}
\end{align}
One has{\footnote{
Note that
$\Phi(x,t) \equiv  \exp(r t^\prime + \epsilon(x^\prime,  t^\prime)) $ is a solution of equation \eqref{bs1m2}
}}
\begin{equation*}
\frac{\partial   \bar{V} }{\partial t^{\prime}}= - \frac{1}{2} \sigma^2 \frac{\partial ^2  \bar{V} }{\partial {x^\prime} ^2}   - \mu   \frac{\partial  \bar{V} }{\partial {x^\prime}}   +  r \, \bar{V} 
\label{bs3mb}
\end{equation*}
\vskip 1mm
Numeraire invariance is analogous to {\it{Phase Invariance}} in quantum mechanics,
However, 
the relevant Black-Scholes gauge group is not the quantum mechanical $U(1)$,  but rather the multiplicative positive real line
$ {\mathbb{R}}^{+}$.
\vskip 1mm
Exponential gauge factors ({\it{cocycles}}) analog to the one in equation \eqref{sch3d}  play a pivotal role in the   {\it{Group Quantization}} formalism.

%% file: group_quantization_formalism.tex
\section{The Group Quantization Formalism}\label{qgroupform}
\vskip 1mm
We describe the main features of  the Group Quantization formalism (\cite{aldaya0}, \cite{wolf11}, \cite{garcia0}, \cite{aldaya22}).
\vskip 1mm
\subsection{The Quantization Group}\label{fiber}
\vskip 1mm
A {\it{Quantization Group}} $\tilde{G}$ is a connected Lie group {\footnote{A Lie group is a differential manifold $G$ endowed with a group composition law $F: G \times G \rightarrow G$ such as $F$ and its inverse are smooth, differentiable applications. }} $\tilde{G}$ which is 
a central  extension of a Lie group $G$ (called the {\it{dynamical group}}) by another Lie group  $U$ called the structural group.
\vskip 1mm
The central extension 
defines a {\it{principal bundle}} $(\tilde{G}, G, \pi, U)$.  The structure group acts on the fiber by left multiplication.  $G$ is called {\it{ base manifold}} of the bundle and  $U$ is called the {\it{ fiber}}.
The projection $\pi$ is a continuous and surjective map $\pi: \tilde{G} \rightarrow G$
\vskip 1mm
\begin{figure}[h]
\includegraphics[width=0.8\linewidth]{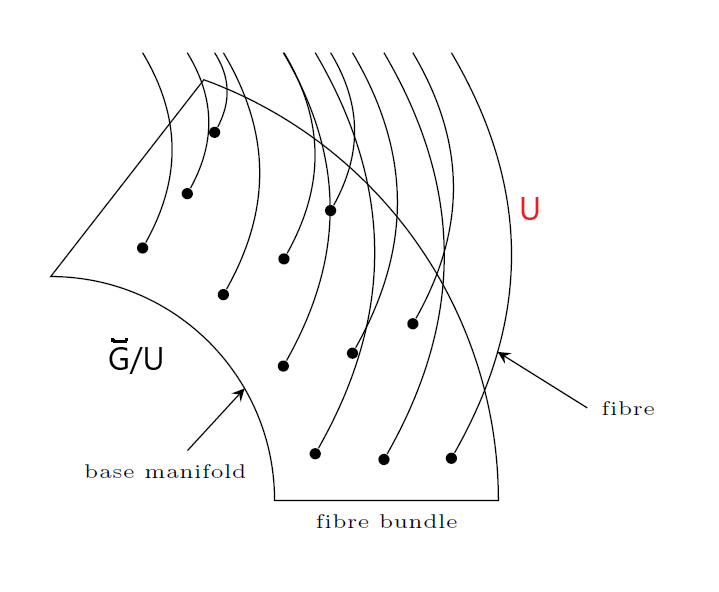} 
\caption{ One can visualize  $\tilde{G}$  by imagining that through each point in the base manifold $G$ there is a line (fiber) of points with different values of the fiber coordinate $\zeta \in U$.}
\centering
\end{figure}
\vskip 1mm
A Quantization Group ${\tilde{G}}$ with structural group $U$ is a Cartan geometry. The Cartan geometry is the geometry of spaces that are locally (infinitesimally) like quotient spaces ${G = \tilde{G}}/U$.
Although G is not a subgroup of   $\tilde{G}$ , $\tilde{G}/U  \simeq G$ as topological spaces any
$\tilde{g} \in  \tilde{G}$ can be decomposed in two parts, $\tilde{g} = (g, u), \, g \in G, u \in U$.   This means  that, locally, $ \tilde{G}$ looks  like the Cartesian product of $G$ and $U$.
\vskip 1mm
In this paper, we will only consider Quantization Groups with structural group $U = {\mathbb{R}}^{+}$. 
\vskip 1mm
\subsection{Invariant Vector Fields}\label{invariantl}
\vskip 1mm
The Group Quantization method considers the action of the group on itself, as opposed to considering the group acting on an external manifold.  This guarantees the existence of two sets of commuting group operators, the right invariant fields and the left invariant fields.
\vskip 1mm
The {\it{left}} and {\it{right}} translations, $L_g$ and $R_{g}$ are defined as
\begin{align}
L_g:G \rightarrow G & \qquad / \quad L_g(g^\prime) = g g^\prime  \nonumber \\
R_g:G \rightarrow G & \qquad / \quad R_g(g^\prime) = g^\prime g
\label{ff2}
\end{align}
$L_g$ and $R_{g}$ are diffeomorphisms of G.  Note that $L_g$ and $R_{g}$ commute.
\vskip 1mm
Let $g,g^\prime,  g^{\prime \prime} \in G $. In an abuse of notation, we write  $g^{\prime \prime} = g^{\prime} g$  for the 
base group $G$ composition law. We denote an element $\tilde{g} \in \tilde{G}$ by $\tilde{g}= (g, \zeta )$, with 
$ \zeta$ in the structural group ${\mathbb{R}}^{+} $. 
The composition law for $\tilde{G}$ is
\begin{equation}
 {\tilde{g}}^{\prime\prime} \equiv (g^{\prime \prime} , \zeta^{\prime \prime} )= (g^{\prime} g, \zeta^{\prime} \zeta \exp(\epsilon(g^\prime, g))
\label{law0}
\end{equation}
where $G\in G$, $\zeta \in  {\mathbb{R}}^{+}$ and $\epsilon(g^\prime, g) \in {\mathbb{R}}$ is the extension cocycle.
\vskip 1mm
A vector field $X$ is called  a left invariant vector field (LIVF) if
\begin{equation*}
{L_{\tilde{g}}^{T}} X_{{\tilde{g}}^\prime} = X_{{\tilde{g}} {\tilde{g}}^\prime}
\label{ff3}
\end{equation*}
and a  right invariant vector field (RIVF) if
\begin{equation*}
{R_{\tilde{g}}^{T}} X_{{\tilde{g}}^\prime} = X_{ {\tilde{g}}^\prime {\tilde{g}}}
\label{ff4}
\end{equation*}
\vskip 1mm
In local coordinates,  the LIVF are given by 
\begin{equation}
{X_i^L}  = \sum_k\,{X_{i,k}^L} \frac{\partial}{\partial g_k}  + \lambda_i \, \Xi 
\label{ivf1}
\end{equation}
with
\begin{align}
& \Xi = \zeta \frac{\partial}{\partial \zeta} \qquad  \qquad \lambda_i \equiv  \left.  \frac{\partial \epsilon(g^\prime, g)}{\partial g_i} \right \rvert_{g = g^\prime, g^\prime=e}  \\ \nonumber
&  {X_{i,k}^L} \equiv \left.  \frac{\partial({g^\prime g)}_k}{\partial g_i} \right \rvert_{g = g^\prime, g^\prime=e} 
\label{ivf2}
\end{align}
\vskip 1mm
and the RIVFs 
\begin{equation}
{X_i^R}  = \sum_k\,{X_{i,k}^R} \frac{\partial}{\partial g_k}  + \gamma_i \, \Xi 
\label{rvf1a}
\end{equation}
where
\begin{align}
& \Xi = \zeta \frac{\partial}{\partial \zeta} \qquad  \qquad \gamma_i \equiv  \left.  \frac{\partial \epsilon(g^\prime, g)}{\partial g_i} \right \rvert_{g^\prime= g, g=e} \\ \nonumber
&  {X_{i,k}^L} \equiv \left.  \frac{\partial({g^\prime g)}_k}{\partial g^{\prime}_i }\right \rvert_{g^\prime= g, g=e}
\label{rvf2a}
\end{align}
The i-th component of a LIVF ${X_i^L}$ is calculated as the derivative respect to {\it{unprimed}} coordinates evaluated at the identity: first set $g^\prime = g$ in the derivative, then set $g=e$.  The i-th component of a LIVF ${X_i^L}$ is calculated as the derivative respect to {\it{primed}} coordinates evaluated at the identity: first set $g = g^\prime$ in the derivative, then set $g^\prime=e$. 
\vskip 1mm
Note that, by construction, RIVFs and RIVFs commute.
\vskip 1mm
\subsection{Cocycles}\label{cocycle}
The function $\epsilon$  in equation \eqref{law0} is called a cocycle. Cocycles are restricted by the group law
properties.
\vskip 1mm
Associativity of the group law
implies that $\epsilon$  must satisfy the following functional equation
\begin{equation}
\epsilon(g^{\prime \prime}, g^\prime) + \epsilon(g^{\prime \prime}   g^\prime, g) = \epsilon(g^{\prime \prime},g^\prime g ) + \epsilon( g^\prime, g) 
\label{extension4}
\end{equation}
and existence of an inverse element
\begin{equation}
\epsilon(g ,e) =  \epsilon(e,g) =  \epsilon(e,e ) =0 
\label{extension4b}
\end{equation}
where $e$ is the identity element of $G$.
\vskip 1mm
\subsection{Coboundaries}\label{coboundaries1}
\vskip 1mm
A coordinate change in the fiber $U$ generates a mathematically trivial cocycle. 
If the fiber elements $\zeta$ undergo the coordinate change $\zeta \rightarrow\zeta\, e^{f(g)}$  in the group law \eqref{law0}, the extension cocycle $\epsilon$ gets an extra additive factor $\delta_c(f)$
\begin{equation*}
\epsilon(g, g^\prime) \rightarrow  \epsilon(g, g^\prime)+ \delta_c(f)(g,g^\prime)  
\end{equation*}
with
\begin{equation}
 \delta_c(f)(g,g^\prime)  = f(g g^\prime) - f(g) -f(g^\prime)
\end{equation}
Trivial cocycles are called {\it{coboundaries}}{\footnote{Not all coboundaries are generated by a coordinate change.}}. The coboundary $\delta_c(f)(g,g^\prime)$ can can be undone by a change of coordinates. 
\vskip 1mm
Although group extensions whose cocycles differ in a coboundary mathematically define the same group
representation,  the dynamical effects generated by coboundaries are not necessarily  trivial.
\vskip 1mm
One useful analogy is a change of variables in a differential equation: differential equations have {\it{canonical forms}} to which they can be reduced,  however it is often convenient to work in 
non-canonical  coordinates. For instance, even if the Black-Scholes equation can be reduced to the simpler heat equation,  in practice its solution is calculated  using variables  which allow to implement naturally the instrument's pricing boundary conditions.
\vskip 1mm
\subsection{Connection. Vertical Form}\label{canonical}
\vskip 1mm
A connection on $\tilde{G}$ is a smooth choice of {\it{horizontal}} subspace $H$ and {\it{vertical}} subspace $V$
such that  $\tilde{G}$ can be decomposed  as a direct sum,  $\tilde{G} = H \oplus V$.
\vskip 1mm
\begin{figure}[h]
\includegraphics[width=0.8\linewidth]{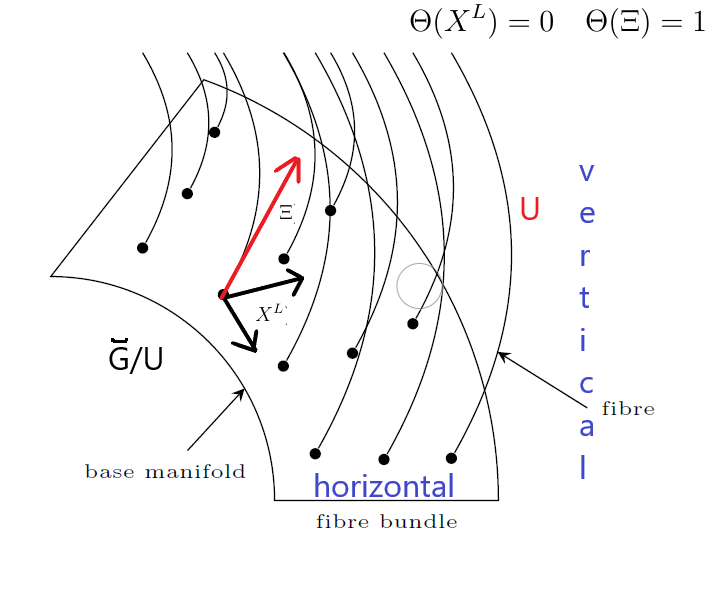} 
\caption{Transformation along the fiber $U$ are {\it{vertical}},  quantities defined on
the base manifold  $G \simeq \tilde{G}/U$  are {\it{horizontal}}.}
\centering
\end{figure}
\vskip 1mm
The  Group Quantization formalism uses  a  Cartan{\footnote{The notion of an affine connection on a differential manifold  ( {\it{Poincar\'e-Cartan}} form ) was introduced by the French mathematician \'Ellie Cartan in two papers
published in 1923 and 1924  \cite{cartan}.}}
connection,
where  the geodesics coincide with the flows of an integrable 
LIVF algebra.  Concretely, 
the connection form in the   Group Quantization formalism  is 
the {\it{vertical}} part $\Theta$ of the {\it{Maurer-Cartan}} form{\footnote{The {\it{Maurer-Cartan}} form. 
 $\Gamma :\tilde{G} \rightarrow {\mathfrak{\tilde{G}}}$  takes values in the Lie algebra of $\tilde{G}$, ${\mathfrak{\tilde{G}}}$
and it  is the unique left-invariant 1-form  such that $\Gamma\rvert_e$  is the identity map. 

 $\Gamma$ can be written 
in terms of the left invariant fields at the identity ${X_i^L}, i = 1,2 \ldots dim(\tilde{G})$ and their dual  forms $\theta$
\begin{equation*}
\Gamma =\sum _{i}{X_i^L}\otimes \theta ^{i}  \qquad  \theta ^{i}({X_i^L}) = \delta_{i,j}
\end{equation*}
The {\it{ Maurer-Cartan equations}} 
\begin{equation*}
 d\theta ^{i}=-{\frac{1}{2}}\sum _{{jk}}{c^{i}}_{{jk}} \, \theta ^{j}\wedge \theta ^{k}.
\end{equation*}
express the differential of the form in terms of the ${c^{i}}_{{jk}}$,  the structure constants of the Lie algebra $ {\mathfrak{\tilde{G}}}$. }}
\vskip 1mm
In this work, the vertical space is the structural group $U= {\mathbb{R}}^{+} $ with a unique generator $\Xi = \zeta \frac{\partial}{\partial \zeta}$ (see equation \eqref{law0}).
\vskip 1mm
 $\Theta$  provides a natural definition of horizontality
by requiring that the horizontal fields belong to the kernel of $\Theta$. 
\begin{equation}
i_{X^L}\, \Theta \equiv \Theta(X^L) = 0  \qquad   i_\Xi \, \Theta  \equiv \Theta(\Xi) = 1   
\label{extension7b}
\end{equation}
where    $X^L$ stands for all LIVFs (Left Invariant Vector Fields).
\vskip 1mm
In coordinates, equations \eqref{extension7b} read (see section \ref{invariantl})
\begin{equation}
\Theta  = \sum _{i}  \, \theta_i \, d g_i + d\, \Xi   \quad \qquad d\, \Xi  \equiv\frac{1}{ \zeta} d \zeta 
\label{extension8}
\end{equation}
where the $\theta_i$ are solutions of a linear algebraic system
\begin{equation}
 \sum _{k}  \, \theta_k \, {X_{i,k}^L}  + \lambda_i = 0  
\label{extension8}
\end{equation}
\vskip 1mm
%
\subsection{Characteristic Module}\label{charmod}
\vskip 1mm
In the Group Quantization formalism, the elements
of the characteristic module $\mathcal{C}_\Theta$  are interpreted as evolution operators.
 $\mathcal{C}_\Theta$ is an integrable system analogous to Hamiltonian fields in classical mechanics,
and 
the integral flows of the $\mathcal{C}_\Theta$ elements are the geodesics of the Cartan connection.
$\mathcal{C}_\Theta$ is defined 
as the set of LIVFs that leave the connection form (vertical form) $\Theta$ strictly invariant
{\footnote{A vector field X is a symmetry of a 1-form  $\Gamma$ if X leaves $\Gamma$ semi-invariant, that is, the Lie derivative
of $\Gamma$ with respect to X is a total differential
\begin{equation*}
L_X \Gamma= d f^{X} \quad  \longrightarrow   \quad d(i_X \Gamma) + i_X (d \Gamma) =   d f^{X}
\label{lie5}
\end{equation*}
where $f^X$ is some function associated with the vector field X. Then, $\mathcal{C}_\Theta = \ker \Theta  \cap  \ker d \Theta $. 
}}
\begin{equation}
X \in \mathcal{C}_\Theta  \quad  \longrightarrow  \quad d(i_X \Theta)  = 0 \quad  i_X (d \Theta) =   0
\label{mod1bb}
\end{equation}
The curvature form $\omega = d \Theta$ can be written in local coordinates as
\begin{align}
\omega & = \sum _{i,j}  \, \frac{\partial  \theta_i }{\partial g_j} d g_j \wedge d g_i = \\ \nonumber 
& \qquad \sum _{i,j} \,  (\frac{\partial  \theta_i }{\partial g_j}-\frac{\partial  \theta_j}{\partial g_i}) dg_j \otimes dg_i
\label{extension7bb}
\end{align}
then, the second of equations \eqref{mod1bb} reads, using the results in section \ref{canonical}
\begin{equation}
\sum _{j} \,  (\frac{\partial  \theta_i }{\partial g_j}-\frac{\partial  \theta_j}{\partial g_i}) \, {X_j} =0
\label{mod1bb2}
\end{equation}
Note that from the Maurer-Cartan equations, $ \omega(X,Y) = 0$ implies $\Theta([X,Y]) = 0$.
\vskip 1mm
\subsection{Polarization Algebra}{\label{FPO}}
\vskip 1mm
The Group Quantization
formalism constructs an irreducible representation of  $\tilde{G}$ starting from the ring ${\mathcal{F}}$ of functions
with domain $\tilde{G}$  and range ${\mathbb{R}}$. 
\vskip 1mm
The reduction is achieved  by restricting the arguments of the functional space  ${\mathcal{F}}$
using a set of horizontal operators and building a polarized function space  ${\mathcal{F_P}}$ that
provides a representation of the group.
\vskip 1mm
The action of vertical fields (generators of the structural group $U$)   $ X^V$ on ${\mathcal{F_P}}$ is
\begin{equation*}
 X^V \Psi   =   f(X^V) \Psi \qquad \forall \, \Psi  \in  {\mathcal{F_P}}
\end{equation*}
where $f(X^V)$ is a function associated to the generator $X^V$. We say that $ {\mathcal{F_P}}$ is a
 $U$- invariant functional space. In this document, $U = \mathbb{R}^{+}$ and $f(X^V) = 1$.
 \vskip 1mm
We will see later in this document that, for the Black-Scholes and the Ho-Lee groups, 
polarized  functions have the meaning of
financial instrument prices.
\vskip 1mm
\subsubsection{First Order Polarization}{\label{FirstPO}}
\vskip 1mm
The  Group Quantization formalism provides a 
natural choice for polarization  by
requiring  invariance respect to
a {\it{horizontal}} algebra $\mathcal{P}$  ( {\it{polarization}} algebra ) 
that imposes constraints on  ${\mathcal{F}}$. 
\vskip 1mm
A {\it{first-order polarization}} (or just {\it{polarization}})  $\mathcal{P}$  is defined as a maximal
horizontal commutative  left  invariant algebra containing the characteristic module $\mathcal{C}_\Theta$, so that 
the constraints are compatible with the evolution operators.
\vskip 1mm
 The
{\it{polarized}} functions ${\mathcal{F_P}}$   will be characterized by conditions of the form
\begin{equation*}
{\mathcal{F_P}} = \{ \Psi  \in  {\mathcal{F}} / X^P \Psi = 0 \quad \forall X^P \in \mathcal{P}  \}
\label{theta55}
\end{equation*}
\vskip 1mm
Since the elements of the algebra $\mathcal{P}$ are integrable vector fields,  a first order polarization defines a {\it{foliation}} of $\tilde{G}$.
This means that it is possible to select functional subspaces on 
on $\tilde{G}$  by requiring them to be constant along integral leaves
of the foliation. 
\vskip 1mm
 \subsubsection{Higher Order Polarization}{\label{SecondPO}}
 \vskip 1mm
When a first order polarization is not able to provide the required functional constraints,
a {\it{higher order polarization}} can be used.
\vskip 1mm
As opposed to first-order polarizations described in (\ref{FirstPO}),
higher-order polarizations  \cite{aldaya6}  contain higher-order differential operators belonging to the left enveloping algebra.
Higher order polarizations do not define a {\it{foliation}} of $\tilde{G}$.
\vskip 1mm
A higher-order polarization  $\mathcal{P_H}$
is a maximal subalgebra of the left-invariant  enveloping algebra  that has no intersection with the generators of the 
structural group $U$ and commutes with the first order polarization  $\mathcal{P}$. 
\vskip 1mm
Commuting with the first order polarization ensures
compatibility with the action of the dynamical operators (RIVFs). As we shall see in the next sections, higher order polarizations can be constructed by using a Casimir operator.
\vskip 1mm
\subsection{Operators}\label{operators}.
\vskip 1mm
Commutativity of the right and left generators makes the RIVFs good candidates for
{\it{quantum operators}}, that is, operators that can reduce the representation from phase space, with coordinates and momenta, to an
irreducible group representation with only coordinates or momenta.
\vskip 1mm
Note that, since $\mathcal{P}$ is spanned by LIVFs, if $X^R$ is a right generator and 	$\Phi$
is polarized
\begin{equation*}
X^P(X^R \Phi) =  X^R(X^P \Phi) = 0  \qquad \forall X^P \in \mathcal{P}   \quad \forall \Phi \in {\mathcal{F_P}}
\label{mod1bb5}
\end{equation*}
so the action of the RIVFs on the space of polarized functions is well defined. Hence
the quantum operators are the restriction of the RIVFs to the polarized $U$- invariant functional space.
\vskip 1mm
\subsection{Noether's theorem}
\vskip 1mm
One can easily verify that the conection $\Theta$ is right invariant, $\Theta(X^R) = 0$ for all RIVFs.
The inner product of the vertical form $\Theta$ and the RIVFs gives the classical Hamiltonian constants of motion.
\vskip 1mm
\subsection{Lagrangian}
\vskip 1mm
The classical Lagrangian  ${\mathcal{L}} (x, \dot{x})$ is obtained by the projection onto the base manifold $G$ of the connection form $\Theta$ along the
 $C_\Theta$  flows (trajectories). See appendix \ref{CLASSA} for a discussion of the {\it{Poincar{\'e}-Cartan}} form of classical mechanics and the Lagrangian function.
\vskip 1mm

%% file: quadratic_sl2R.tex
\section{The  $WSp(2,  \mathbb{R} )$ Group}\label{SL2RGROUP}
\vskip 1mm
\subsection{Heisenberg-Weyl Group.}\label{HWGROUP}
\vskip 1mm
The  Heisenberg-Weyl group $W$ 
 is a central extension of the two dimensional Euclidean translation  group. Its algebra  
is generated by three elements, $\mathds{ P} $,  $\mathds{ Q} $  and $\mathds{ 1}$, where $\mathds{ 1}$ is the identity operator. The 
only non trivial commutator is 
\begin{equation}
\label{weyl1}
[\mathds{ P},\mathds{ Q}] = \gamma \,  \, \mathds{1} \qquad   \qquad \gamma \,  \in \mathbb{ C}
\end{equation}
It can be proven that the choices for  $\gamma$  are equivalent to select  $\gamma$ real or $\gamma$ pure imaginary.
\vskip 1mm
Let $p, q \in \mathbb{R}$  , $\theta \in \mathbb{C}$ and define
\begin{equation}
\label{weyl2}
\mathds{W}(\theta,x,q)  \equiv    \exp( \theta \mathds{1}+ x \, \mathds{ P}  +  p \mathds{ Q}) 
\end{equation}
The operators $\mathds{W}$   
specify{\footnote{The action of $W$ on  $sl(2,\mathbb{R}$ is
\begin{equation}
\mathds{W} (a\, \mathds{Q} + b \, \mathds{P} + \eta \, \mathds{I}) \mathds{W}^{-1} = a\, \mathds{Q} +  b \mathds{P} + (\eta \, +  w (a \, p - b\, x) ) \mathds{1}
\end{equation}
therefore, $W$ acts on ${\mathfrak{w}}$ as a three-dimensional space with
Cartesian coordinates $(p, x, \theta)$.}} a group composition law under multiplication{\footnote{ Exponentiation and composition of operators are to be considered in the context of formal operator series.}}. 
\begin{equation}
\label{mult2}
\mathds{W}^{\prime \prime} ( p^{\prime \prime}, x^{\prime \prime},\theta^{\prime \prime})= \mathds{W}^\prime (  p^{\prime}, x^{\prime},\theta^{\prime}) \, \mathds{ W} (  p, x,\theta)
\end{equation}
Using the  the   {\it{Baker-Campbell-Hausdorff}} formula{\footnote{The {\it{Baker-Campbell-Hausdorff}} formula reads
\begin{align}
\label{weyl3b}
 \ln(e^X e^Y) & = X + Y + \frac{1}{2} [X,Y] +  \nonumber \\
               &  \frac{1}{12} ( [X,[X,Y]] - [Y,[X,Y]]) + \ldots 
\end{align}
when  $[X,Y]$ is a number 
\begin{equation}
\label{weyl2}
e^X e^Y  = e^{X + Y + \frac{1}{2} [X,Y]}   \qquad \rightarrow \quad  e^X e^Y  = e^Y e^X e^{ [X,Y]}
\end{equation}
One can prove that
\begin{equation}
\label{weyl3}
e^X e^Y  e^{-X} =  e^X Y e^{-X }= Y + [X,Y] +  \frac{1}{2}[X, [X,Y]] + \ldots
\end{equation}
}}
, we obtain
\begin{align}
p^{\prime \prime} & = p +  p^\prime \nonumber \\
x^{\prime \prime} & =  x+  x^\prime  \nonumber \\
\theta^{\prime \prime} & =  \theta + \theta^\prime   +  \frac{1}{2} \gamma \,   (p x^\prime  - x p^\prime  ) \label{weyl3}
\end{align}
 \vskip 1mm
The group law \eqref{weyl3} provides a coordinate representation{\footnote{These are {\it{right invariant}} generators, corresponding to the action of $\mathds{W}^\prime $.}} of the abstract operators  $\mathds{ P}$, $\mathds{ Q}$   and $\mathds{1}$  acting on the functional space
\begin{equation}
\Psi(p , q , \theta) = \exp(\theta)\, \psi(p,q)
\end{equation}
that provides the required Heisenberg-Weyl commutation relations
\begin{align}
\mathds{ P} \mapsto X_p & = \frac{\partial}{\partial p} -  \frac{1}{2} \gamma \,  x \,  \frac{\partial}{\partial \theta} \nonumber \\
\mathds{Q} \mapsto X_x  & = \frac{\partial}{\partial x} +  \frac{1}{2} \gamma \,  p\,  \frac{\partial}{\partial \theta}  \nonumber \\
\mathds{1} \mapsto  X_\theta & = \frac{\partial}{\partial \theta}  
\end{align}
We get 
\begin{equation}
\qquad [ X_p, X_x] = \gamma \,   X_\theta 
\end{equation}
\vskip 1mm
\subsection{The  $SL(2, \mathbb{R})$  group}
\vskip 1mm
$ SL(2, \mathbb{R}) $, the special (unimodular) linear group in two real dimensions, 
has a natural representation  by $2 \times 2$ real matrices with determinant one
\begin{equation}
\label{amatrix}
{\bf{M}} = \begin{pmatrix}
    a & b \\
    c & d 
\end{pmatrix}  \in SL(2, \mathbb{R}) , \quad  a d - b c = 1  
\end{equation}
\vskip 1mm
The $SL(2, \mathbb{R})$ group is an automorphism of the $W$ group,  preserving  the algebra commutators.  Define new operators $\mathds{P}^\prime$ and $\mathds{Q}^\prime$
as linear combinations of  $\mathds{P}$ and $\mathds{Q}$, using  the matrix  \eqref{amatrix}
\begin{align}
 \mathds{P}^{\prime} = & a \, \mathds{P} + b\,  \mathds{Q} \nonumber \\
 \mathds{Q}^{\prime} = & c  \, \mathds{P}  + d \,  \mathds{Q}
\end{align}
then, by direct computation,
\begin{equation}
[\mathds{P}^{\prime}, \mathds{Q}^{\prime}] = [\mathds{P}, \mathds{Q}]
\label{auto1}
\end{equation}
\vskip 1mm
$ SL(2, \mathbb{R})$ belongs to a class of Lie groups called the symplectic groups  $ Sp(2n, \mathbb{R})$
which leave invariant a skew-symmetric form and  play
an important role in the geometry of phase space and Hamiltonian systems.
One can verify the isomorphism $SL(2,  \mathbb{R}) \approx Sp(2, \mathbb{R})$ for the two-dimensional symplectic matrix $ {\bf{\mathrm{\Omega \, }}}$
\begin{equation}
{\bf{M^\top}} {\bf{\mathrm{\Omega \, }}}  {\bf{M}} = {\bf{\mathrm{\Omega \, }}} \qquad \qquad {{\bf{\mathrm{\Omega \, }}}} \equiv \begin{bmatrix}
    0 & 1 \\
    -1 & 0 
\end{bmatrix}
\label{omegadef}
\end{equation}
\vskip 1mm
\subsection{The $WSp(2,  \mathbb{R} )$ group}
\vskip 1mm
$WSp(2,  \mathbb{R} )$  is a real, non-compact, connected, simple Lie group  that is the semi direct product of the two-dimensional
real symplectic group  $Sp(2,  \mathbb{R})$($\approx SL(2,  \mathbb{R})$) by the Heisenberg-Weyl group $W$. 
\vskip 1mm
The 
$WSp(2,  \mathbb{R} )$  group is a subgroup of  the Schrodinger group in one dimension, and is called{\footnote{Our composition law differs from  \cite{wolf2} in that the extension group in  \eqref{slr2a} is $\mathbb{R}^{+}$, not  $\mathbb{R}$.}}
 {\it{Group of Inhomogeneous
Linear Transformations}} in reference \cite{wolf2}. 
\vskip 1mm
Let  an element $g \in WSp(2,  \mathbb{R} )$ be parametrized by $g \left( {\bf{M}}, \bf{u}, \zeta \right)$ 
where $ {\bf{M}} \in SL(2, \mathbb{R})$, $ \zeta  \in  {\mathbb{R}}^{+}$ and
${\bf{u}} \equiv (p,x)  \in  {\mathbb{R}}^2 $.
The $WSp(2,  \mathbb{R} )$  composition law  is
\begin{equation}
g^{\prime\prime} \left( {\bf{M}}^{\prime\prime}, \bf{u}^{\prime\prime}, \zeta^{\prime\prime} \right) =  g^{\prime} \left( {\bf{M}}^{\prime}, \bf{u}^{\prime}, \zeta^{\prime} \right) \, g \left( {\bf{M}}, \bf{u}, \zeta \right)
\end{equation}
with
\begin{align}
{\bf{M}}^{\prime \prime} & = {\bf{M}}^{\prime } {\bf{M}}   \nonumber \\
 \qquad \bf{u}^{\prime \prime}  &  = \bf{u}^{\prime } {\bf{M}} +    \bf{u} \nonumber \\
\zeta^{\prime \prime} & = \zeta^\prime \, \zeta \, \exp \left(-  \frac{1}{2} \gamma \, \bf{u}^{\prime } {\bf{M}} {\bf{\mathrm{\Omega \, }}}  \bf{u}^{T}   \right) 
\label{slr2a}
 \end{align}
where ${\bf{\mathrm{\Omega}}}$ is the two-dimensional symplectic matrix  in \eqref{omegadef} and
 $\gamma$  is given by  the $[\mathds{ P},\mathds{ Q}]$ Heisenberg-Weyl commutator in
\eqref{weyl1}. 
\vskip 1mm
\subsubsection{Weyl Commutation Relations}
\vskip 1mm
In the mathematics literature, such as \cite{wolf2} and \cite{Miller}, the usual convention is that the momentum  and position operator commutator is $[\mathds{ P},\mathds{ Q}] = - i $,
which corresponds to an extension{\footnote{In quantum mechanics the cocycle in \eqref{slr2a} is multiplied 
by a factor $1/\hslash$ for dimensionality reasons, and   $[\mathds{ P},\mathds{ Q}] = - i \, \hslash$.}} of the 2-dimensional translations by $U(1)$. 
Then, for diffusive equations such as the heat equation, one makes the theory  {\it{Euclidean}} at the last step by setting  the time variable $t$ to $i \, t$. 
\vskip 1mm
In this work  we set $\gamma \, = 1$, corresponding to a central extension of the Euclidean translations by $\mathbb{R}^{+}$.  The reason is  
that in finance we want $\mathds{ P}$ to represent a change (a delta)  in the quantity $\mathds{ Q}$, wich usually represents a a price or a rate. This choice has the added advantage that  there is 
no need to consider a passage to a fictitious  {\it{Euclidean}} time, since the time variable in the group represents the actual calculation time.
\vskip 1mm
When we quote results from  to the mathematics literature  (for instance in section \ref{WST}) we will assume,
unless otherwise indicated, that the results are obtained with the usual commutator $[\mathds{ P},\mathds{ Q}] = - i$.
\vskip 1mm
\subsubsection{Orbits and Subgroups}
\vskip 1mm
The 
$WSp(2,  \mathbb{R} )$  generates the dynamics for quadratic  Hamiltonians (\cite{Miller2}, \cite{Miller}, \cite{wolf11}, \cite{wolf2} ) by the action of higher order operators on $W$.
The adjoint action 
of the group consist of six distinct orbits  \cite{wolf2}. Representative of these orbits are given by the following operators, that include quadratic elements from the  enveloping algebra of  $W$
\begin{equation*}
 \mathds{P}^2 \, , \,   \mathds{ P}^2  + \mathds{Q} \, , \,  \mathds{ P}^2  + \mathds{Q}^2 \, , \, \mathds{ P}^2  - \mathds{Q}^2 \, , \, \mathds{P} \, , \, \mathds{1}
 \end{equation*}
 \vskip 1mm
We have the following isomorphisms: $g \left( {\bf{M}}, {\bf{0}}, 1 \right) \approx SL(2, \mathbb{R})$  and $g \left( {\bf{1}}, \bf{u}, \zeta \right) \approx W$.
The unit element is $g \left( {\bf{1}}, {\bf{0}}, 1 \right)$
and the inverse element is $g \left( {\bf{M}}^{-1}, -{\bf{u}}\, { {\bf{M}}^{-1}},  {1 / \zeta} \right)$
\vskip 1mm
The different quantization groups considered in this paper will be distinguished subroups of $WSp(2,  \mathbb{R} )$.
\vskip 1mm
In physics, the 
$WSp(2,  \mathbb{R} )$  includes as subgroups the symmetry group of the free particle, the gravitational free-fall, as well as
the symmetry group of the ordinary harmonic oscillator and the {\it{repulsive}}
harmonic oscillator (with imaginary frequency). 
In finance, we will obtain the Black-Scholes theory, 
the Ho-Lee model and the Euclidean attractive and repulsive oscillators.  


\vskip 1mm
\subsection{Linear Canonical Transformations}\label{WST}
\vskip 1mm
$WSp(2,  \mathbb{R} )$ acts on functional spaces as
integral transforms{\footnote{These transforms also define pseudo-differential (hyperdifferential) operators  associated to the corresponding  $SL(2, \mathbb{R})$ kernel  \cite{wolf2}.}} 
called  {\it{Linear Canonical Transformations}},
whose kernel  is a  $SL(2, \mathbb{R})$ matrix  \cite{wolf2}
 \begin{equation}
g \left( {\bf{M}}, \bf{u} \right) (f(x)) = \int_{ \mathbb{R} } W({\bf{M}}, x,x^\prime) f(x^\prime) d x^\prime
\label{trans1}
 \end{equation}
with kernel (using the notation of equation \eqref{amatrix})
\begin{equation}
 W({\bf{M}},x,x^\prime)  = \frac{e^{-i \pi/4}}{\sqrt{2 \pi b}}\, \exp( i ( a {x^\prime}^2 - 2 x x^\prime + d\, x^2)/(2 \,b))
 \label{ltcdef}
 \end{equation}
\vskip 1mm
Linear Canonical Transformations (LICs),
have the important property that composition of the transforms is equivalent  to 
multiplication of their  $SL(2, \mathbb{R})$  kernels (\cite{wolf11}, \cite{wolf2}). 
\vskip 1mm
LICs  kernels  can be analytically continued to $SL(2, \mathbb{C})$,
subject to some restrictions  \cite{wolf11}. In this work, we will use, when indicated,  equation \eqref{trans1} with the mapping $b \rightarrow - i b$
\begin{equation}
 W({\bf{M}},x,x^\prime)  = \frac{1}{\sqrt{2 \pi b}}\, \exp(  - ( a {x^\prime}^2 - 2 x x^\prime + d\, x^2)/(2 \,b))
 \label{ltcdef2}
 \end{equation}
which has been modified from  \eqref{ltcdef} because our choice for the  Heisenberg-Weyl commutator $[\mathds{ P},\mathds{ Q}] = 1 $.
\vskip 1mm
The Fourier transform, the Laplace transform,  the Bargmann transform and the Mellin transform are examples of LICs. 
\vskip 1mm
\subsubsection{Fourier Transform}\label{FT}
The  Fourier transform 
\begin{equation*}
\label{lm0}
  {\mathcal {F}} ( f)(z) = \int _{-\infty}^{\infty } f(p) \, e^{i p z} dp 
\end{equation*}
has the $SL(2, \mathbb{R})$ kernel
\begin{equation*}
 {\bf{F}} = \begin{pmatrix*}[c]
     \phantom{-}0 & 1 \\
    -1 & 0 
\end{pmatrix*}
\end{equation*}
\subsubsection{Double Sided Laplace Transform}\label{LMT}
The double sided Laplace  transform 
\begin{equation}
\label{lm0}
  {\mathcal {L}} ( f)(z)  = \int _{-\infty}^{\infty } f(p) \, e^{ p z} dp 
\end{equation}
can be obtained from $ {\mathcal {F}}$ by the formal map $p \mapsto i p$. Its kernel is a $SL(2, \mathbb{C})$  matrix
\begin{equation*}
 {\bf{L}} = \begin{bmatrix}
    0 & i \\
    i & 0 
\end{bmatrix}
\end{equation*}
\vskip 1mm
In spite of formal similarities, existence of a Fourier transform exists does not automatically imply the existence of a Laplace transform \cite{wolf2}.
\vskip 1mm
The inverse transform is 
\begin{equation}
 {\mathcal {L}}^{-1} ( f)(x)  =   \frac{1 }{2 \pi i} \, \int _{c-i \infty}^{c+ \infty } \, dp \,  e^{ x p} \, f(p) 
\label{price1bb}
\end{equation}
The integration contour is   a
vertical line on the complex plane, such that all singularities of the integrand lie on the left of it.
\vskip 1mm
\subsubsection{Bargmann Transform}\label{BAT}
The Bargmann transform has the complex kernel
\begin{equation*}
 {\bf{B}} = \frac{1}{\sqrt{2}} \, \begin{pmatrix*}[c]
   \phantom{-} 1 & -i \\
    -i &  \phantom{-}1 
\end{pmatrix*}
\end{equation*}
\vskip 1mm
\subsubsection{Mellin Transform}\label{MT}
\vskip 1mm
The Mellin transform ${\mathcal {M}}$  
 \begin{equation}
 {\mathcal {M}} (f) (z)  \equiv  F(z) = \int _{0}^{\infty }y^{-z-1} f(y)\,d y
\label{mellin1}
\end{equation}
 is obtained from the Laplace transform ${\mathcal {L}}$ by the change of coordinates $p \rightarrow - \ln(y)$. 
\vskip 1mm
The resolution of identity
\begin{equation*}
\frac{1}{2 \pi i} \, \int _{c- i \infty}^{c+ i \infty } x^{-s} y^{s} d s= y \delta(x-y)
\end{equation*}
leads to the Mellin inversion formula 
\begin{equation*}
 {\mathcal {M}}^{-1}(f) (y) \equiv f( y)  = \frac{1}{2 \pi i} \, \int _{c-i \infty}^{c+ i \infty } y^{z} F (z) \,dz
\end{equation*}
\vskip 1mm  
The Mellin transform has  been more widely used in finance
than the double sided Laplace transform  (\cite{Mellin0}, \cite{Mellin2}, \cite{Mellin4},  \cite{Mellin5}). 
The sign of the transform variable $z$ in \eqref{mellin1} differs from the usual definition in the mathematics literature in order to
calculate  transforms of positive powers of the stock price.

%% file: black_scholes_nokets.tex
\section{Black-Scholes}\label{BSQUANT}
\vskip 1mm
\subsection{Quantization Group}
\vskip 1mm
We apply the Group Quantization formalism to the 
 $WSp(2, \mathbb{R})$ subgroup generated by the parabolic $SL(2, \mathbb{R})$  matrix
 \begin{equation}
 {\bf{M}}_{BS}= \begin{bmatrix}
    1  &  \,  {\sigma^2} t\\
    0  &  1
\end{bmatrix}  \qquad  \sigma^2 \in {\mathbb{R}} \qquad  t \in {\mathbb{R}}
\label{genmatrixBSa}
\end{equation}
\vskip 1mm
${\bf{M}}_{BS}$ has only one eigenvalue  $\lambda = +1$ and it cannot be diagonalized. 
${\bf{M}}_{BS}$ is a {\it{shear}} transformation on $\mathbb{R}^2$: it leaves the upper plane invariant, while it displaces the lower plane by $ {\sigma^2} t$. This will be significant  
when finding a first order polarization in section \ref{FPO}.
\vskip 1mm
\subsection*{Composition Law}
The $WSp(2, \mathbb{R})$ subgroup generated by the matrix \eqref{genmatrixBSa} constitutes
the Black-Scholes quantization group $\tilde{B}$. Its 
composition law  is obtained from  the   $WSp(2,  \mathbb{R} )$  composition law 
\eqref{slr2a} and the matrix ${\bf{M}}_{BS}$, with some cocycle modifications that we will explain below{\footnote{With no modifications, the Group Quantization formalism  leads to the heat equation.}}
\begin{align}
t^{\prime \prime} &  =t + t{^\prime}             \nonumber \\
p^{\prime \prime} &  = p^{\prime}  + p      \nonumber \\
x^{\prime \prime} & =x^{\prime }  + x + \sigma^2 p^{\prime} t         \nonumber \\
\zeta^{\prime \prime} & =\zeta^{\prime } \zeta e^{   \epsilon_{BS}(g,g^\prime) }   
\label{cocyclebs1a}
\end{align}
where $ t,p,x\in {\mathbb{R}}$ and $\zeta \in {\mathbb{R}}^{+}$ 
When deriving the group law, we have used  that
\begin{equation*}
 {\bf{M}}_{BS}^{\prime \prime}  = {\bf{M}}_{BS}^{\prime}\,  {\bf{M}}_{BS} \quad \Rightarrow  \quad t^{\prime \prime}   =t + t{^\prime} 
\end{equation*}
\vskip 1mm
The composition law for $(t,p,x)$ gives the familiar Galilean transformations: the Galilei group $G$ is the base group of  $\tilde{B}$,  $G \simeq \tilde{B}/ {\mathbb{R}}^{+} $,
with $\sigma^2 p$ corresponding to the Galilean boost..
\vskip 1mm
Our interpretation of the group coordinates is that $x$ represents the (dimensionless) logarithm of the stock price, $S \equiv S_0 \, e^x$,
$p$ is the conjugate momentum of $x$,  $t$ is the calendar time, and
$\sigma$ is the stock  volatility. Notice that interest rates are not explicitly present in the group law. Interest rates in the Black-Scholes theory are not dynamical quantities and they
will be incorporated in section \eqref{IRDEF} as a coordinate change in the fiber $\zeta$.
\vskip 1mm
The extension cocycle $\epsilon_{BS}$ 
is the sum of the true cocycle{\footnote{ All true Galilean cocycles  with the same $\sigma$   
 differ in a coboundary from the Galilean cocycle \eqref{galileancocycle},  i.e., they can be {\it{undone}} by a change of coordinates. 
The  exception occurs only for the one dimensional Galilean group. In one dimension there is  another true cocycle which will be used in section  \ref{IRQUANT}. 
}}  $\epsilon_{G}$  and the coboundary $\epsilon_{N}$.
\begin{equation}
\epsilon_{BS}(g,g^{\prime})  =  \epsilon_{G}(g,g^{\prime})  \, +  \,  \epsilon_{N}(g,g^{\prime})  
 \label{cocyclebs2a}
\end{equation}
For later convenience when defining the Laplace transform, we use as  Galilean cocycle{\footnote{
 $\epsilon_{G}$ equals 
 the $WSp(2,  \mathbb{R} )$ cocycle in \eqref{slr2a} 
plus the coboundary generated by $(p x)/2$   (see section \ref{coboundaries1}).
 \vskip 1mm
 The $WSp(2,  \mathbb{R} )$ cocycle is
\begin{equation}
\epsilon_{W}(g,g^{\prime})   \equiv   - \frac{1}{2} { \bf{u}^{\prime } {\bf{M_{BS}}} {\bf{\mathrm{\Omega}}}  \bf{u}^{T} } =  
 \frac{1}{2}( x p^{\prime}  - p x^{\prime}  +  \sigma^2 \, p  {p^\prime}  t  )
\label{galileancocycle1}
\end{equation}
The coboundary generated by $(p x)/2$  is
\begin{equation}
 \delta_{c} ( \frac{1}{2} \, p x)=  \frac{1}{2} ( x^{\prime} p +  x p^{\prime}  + \sigma^2  {p^{\prime}}^2 t +\sigma^2  p p^{\prime} t ) 
\end{equation} }}
\begin{equation}
\epsilon_{G}(g,g^{\prime})  = p^{\prime} x  +  \sigma^2 \, p {p^\prime}  t  +  \frac{1}{2} \sigma^2 {p^\prime}^2  t   
 \label{galileancocycle}
\end{equation}
\vskip 1mm
$\epsilon_{N}$ represents a numeraire choice{\footnote{This coboundary will provide a constant first derivative coefficient in the Black-Scholes equation.
It is possible  to add a new group coordinate, much like a quantum gauge potential  (see \cite{calixto}), which  allows the introduction of numeraires  that depend on stock price and time. This approach will be explored in a future work.}}
\begin{equation}
\epsilon_{N} (g,g^{\prime}) \equiv \delta_c(f_N)   =   \mu\,  p^{\prime} t   
\label{galileancob} 
\end{equation}
and has the generating function (see section \ref{coboundaries1})
\begin{equation*}
f _N(g) =  \frac{\mu}{\sigma^2} x 
\end{equation*}
\vskip 1mm
\subsection{Stock Volatility} \label{variance}
\vskip 1mm
Bargmann ({\cite{bargmann1}, \cite{bargmann2}) used a central extension of the Galileo group with  $U(1)$
for the description of the free non relativistic quantum mechanical particle.
The cocycle  used by  Bargmann was 
\begin{equation}
 \epsilon_{B} (g,g^{\prime})  = - \frac{i }{2 \hslash } m ( x v^{\prime}  - v x^{\prime}  +  \, v  {v^\prime}  t  )
\label{bargman1}
\end{equation}
which is labeled by the particle mass $m$. Bargmann proved that two cocycles of the form \eqref{bargman1}
with different values of $m$ are not equivalent, they cannot be transformed into each other by a coboundary.
\vskip 1mm
The cocycle \eqref{bargman1} equals the $WSp(2,  \mathbb{R} )$ cocycle
\begin{equation}
  - \frac{1}{2} { \bf{u}^{\prime } {\bf{M_{BS}}} {\bf{\mathrm{\Omega}}}  \bf{u}^{T} } =  
 \frac{1}{2}( x p^{\prime}  - p x^{\prime}  +  \sigma^2 \, p  {p^\prime}  t  )
\label{galileancocycle1}
\end{equation}
with the  substitutions
\begin{equation*}
\sigma^2 \rightarrow i \frac{\hslash}{m} \quad \qquad p \rightarrow \frac{v}{\sigma^2} 
\label{barg1}
\end{equation*}
The cohomological importance of mass in in quantum mechanics has been extensively studied in the literature. The stock volatility plays a similar role in finance. 
Formally, the Black-Scholes theory is the quantum mechanics of a free particle with an imaginary mass. The classical limit $\hslash \to 0$ corresponds to the zero volatility limit.
\vskip 1mm
\subsection{Lie Algebra}{\label{IVF}}
\vskip 1mm
\subsubsection*{Left Invariant Vector Fields}
\begin{align}
 {X_{p}^L} & = \frac{\partial}{\partial p}  + x \, \Xi \qquad \qquad  {X_{x}^L}  = \frac{\partial}{\partial x}   \nonumber \\
{X_{t}^L}  & = \frac{\partial}{\partial t} +  \sigma^2  p \, \frac{\partial}{\partial x}    + E(p) \, \Xi  \nonumber \\
 {X_{\zeta}^L} & \equiv  \Xi = \zeta \frac{\partial}{\partial \zeta}   
\label{leftXbs}
\end{align}
with
\begin{equation}
 \qquad E(p) \equiv \frac{1}{2 } \sigma^2 \,p^2 + \mu p   
 \label{energy}
 \end{equation}
\vskip 1mm
\subsubsection*{Commutators}
The non-zero commutators are
\begin{equation}
[{X_{t}^L} , {X_{p}^L} ]  = - \sigma^2 {X_{x}^L}   - \mu  \Xi  \qquad \quad  [{X_{p}^L} , {X_{x}^L} ] = \, - \Xi  
\label{liebrakets0}
\end{equation}
 \vskip 1mm
\subsection{Connection}
\vskip 1mm
The expression for the 
the vertical form $\Theta$  and the curvature form $ d \Theta $ are (see section \ref{canonical}) 
\begin{align}
& \Theta  =    -x d p  -  E(p) \, dt   + d \, \Xi    \nonumber \\
& d \Theta  =   d p \wedge  d x   -  (\sigma^2  p + \mu) \, dp \wedge dt 
\label{theta2bs}
\end{align}
with
\begin{equation}
 \qquad \qquad d\, \Xi  \equiv\frac{1}{ \zeta} d \zeta 
 \label{energy}
 \end{equation}
\vskip 1mm
\subsubsection{Characteristic Module}\label{charmod}
\vskip 1mm
$\mathcal{C}_\Theta$ 
is  generated by a unique field $X_C $ 
\begin{align}
&X_C  = {X_{t}^L} + \mu  \, {X_{x}^L} = \nonumber \\
&\qquad \frac{\partial}{\partial t} + (\sigma^2 p +   \mu )  \frac{\partial}{\partial x} + E(p)  \, \Xi  
\label{mod2bbbs}
\end{align}
 The restriction of $\Theta$ to the $\mathcal{C}_\Theta$ flows gives the Black-Scholes Lagrangian, see appendix \ref{LAGBS}.
 \vskip 1mm
\subsubsection{Interest Rates}\label{IRDEF}
\vskip 1mm
We introduce interest rates  by performing 
a change of coordinates{\footnote{This is equivalent to a non-horizontal polarization \cite{aldaya22}, with constraints of the form $X \,\Psi = a \Psi, a \in \mathbb{R}$ rather than $X \, \Psi = 0$. }} in 
the fiber $\zeta$. The mapping
\begin{equation*}
\zeta \mapsto \tilde{\zeta} =  \zeta \, e^{ r t}  \qquad r \in \mathbb{R} 
\end{equation*}
does not change the group law  \eqref{cocyclebs1a},   since $t^{\prime \prime}  =t + t{^\prime}$, 
however it
alters the separation between vertical and horizontal spaces. 
Under this coordinate change, the structural group generator $\Xi$ becomes
\begin{equation*}
\Xi =  \zeta \frac{\partial}{\partial \zeta}   \quad  \mapsto \quad  \tilde{\Xi} =    \tilde{\zeta} \frac{\partial}{\partial \tilde{\zeta}}
\end{equation*}
 and it dual  form  $d \Xi$ acquires a horizontal component
\begin{equation}
  d \, \Xi  = \frac{1}{ \zeta} d \zeta  \mapsto  \,  \frac{1}{ \tilde{\zeta}} d \tilde{\zeta} +  +   r \, dt  \equiv  d \,  \tilde{\Xi }    +   r \, dt
  \label{tilda1}
\end{equation}
The vertical connection form $\Theta$  is modified  in a straightforward way by equation \eqref{tilda1}
\begin{equation*}
\Theta \mapsto  \tilde{\Theta} = \Theta  + r \, dt
\end{equation*}
\vskip 1mm
Therefore, the time component ${f_t^L}$ of the LIVFs  (determined by $\tilde{\Theta} ( X^{L}) = 0$)  picks a new vertical part
\begin{equation*}
{f_t^L}\frac{\partial }{\partial t}  \mapsto  {f_t^L}\frac{\partial }{\partial t}  - r \tilde{\Xi }
\end{equation*}
which induces the following change in the left time generator 
\begin{equation*}
{X_{t}^L} \mapsto  {X_{t}^L} - r \tilde{\Xi }  
\end{equation*}
Note that one obtains the same Lie algebra commutators \eqref{liebrakets0} in the new coordinates.  The RIVFs  are not modified by this coordinate change.
\vskip 1mm
In the rest of this section we use the expression of the vectors fields and connection in these new coordinates and we omit the notation $\tilde{\zeta}$ , $\tilde{\Theta}$, etc. for brevity.
\vskip 1mm
\subsection{First Order Polarization}{\label{FPO}
\vskip 1mm
\subsubsection{Black-Scholes Equation in Momentum Space}
\vskip 1mm
In section \ref{BSQUANT} we noted that  ${\bf{M_{BS}}}$ is a shear mapping where only the $p$ space (upper plane) is invariant, and that ${\bf{M_{BS}}}$ cannot be diagonalized. These features 
indicate that the only possible first order polarization is a polarization in $p$-space, as we will verify below.
\vskip 1mm
The first  order polarization algebra $\mathcal{P}$ is spanned
{\footnote{${X_{p}^L}$ cannot belong to    $\mathcal{P}$  because 
\begin{equation*}
[{X_{t}^L} , {X_{p}^L} ]  = - \sigma^2 {X_{x}^L}   - \mu \,  \Xi 
\end{equation*}
}}
 by the $x$-translations and the  $\mathcal{C}_\Theta$ generator $X_C$ 
\begin{equation}
{\mathcal{P}} = \langle {X_{x}^L} , {X_C}  \rangle 
\label{pol0bs}
\end{equation}
\vskip 1mm
The polarized functions $\Psi \in {\mathcal{F_P}}$ are found by imposing the polarization constraints
on the functional space ${\mathcal{F_P}}$
\begin{equation*}
{\mathcal{F_P}} = \{\, \Psi : \tilde{G} \rightarrow \mathbb{R} \, / \,  \Psi(x,p, t,\zeta) = \zeta \Psi(x,p,t)  \, \}
\label{theta5}
\end{equation*}
One has  
\begin{subequations}
\begin{align}
{X_{x}^L} \Psi  = 0 &  \rightarrow    \Psi = \zeta  \, \psi(p,t)  \nonumber  \\
X_C \Psi  = 0 &  \rightarrow   \nonumber  \\
& \frac{\partial \psi}{\partial t}+ \frac{1}{2} \sigma^2 p^2 \, \psi + \mu \, p \, \psi - r \, \psi = 0  \label{fpol1}
\end{align}
\end{subequations}
Equation \eqref{fpol1} is the Black-Scholes equation in momentum space.
\vskip 1mm
\subsubsection{Polarized Functions}
\vskip 1mm
A separable solution of  the Black-Scholes equation \eqref{fpol1} is
\begin{equation*}
 \tilde{\psi} (p, t)=  e^{-E_{r}(p) t} \, \Phi(p)  
\end{equation*}
with  $\Phi(p)$ an arbitrary function of $p$ and
\begin{equation*}
 E_{r}(p) \equiv \frac{1}{2 } \sigma^2 \,p^2 + \mu p -r
 \end{equation*}
We can write a general polarized function as an inverse Laplace transform (a Linear Canonical  Integral transform)
\begin{align}
 \Psi (\zeta, x,  t) & =  \nonumber \\
 & \zeta  \,   \frac{1 }{2 \pi i} \, \int _{c-i \infty}^{c+ \infty } \,    K(p, t)  \exp( p x) \,\Phi(p) \, d p 
\label{pol1bs0}
\end{align}
with 
\begin{equation}
\qquad K(p, t)   = \exp(-  E_{r}( p) \, t) 
\label{momkernel}
\end{equation}
The integration measure $d p$ is  the dual form of ${X_{p}^L}$, the vector field absent from the polarization algebra.  
We will justify the use of the inverse Laplace transform in section \ref{OPX}.
\vskip 1mm
In the Black-Scholes theory, polarized functions represent prices of financial derivatives, whereas 
the $\Phi(p)$ correspond to  terminal (payoff) conditions.  Appendix   \ref{BRAKET} shows how to price financial instruments in momentum space. 
\vskip 1mm
\subsection{High Order Polarization}\label{HOP}
\vskip 1mm
\subsubsection{Black-Scholes Equation in Coordinate Space}
\vskip 1mm
The Black Scholes equation can be obtained directly in the coordinate representation by using a {\it{high-order}} polarization (section {\ref{SecondPO}}).
\vskip 1mm
The second order operator 
\begin{equation}
C_{P}  =   {X_{t}^L} + \mu   {X_{x}^L}  +  \frac{1}{2}\,\sigma^2 \,  X_{x}^{L} \,X_{x}^{L}  
\label{cas1bs}
\end{equation}
is a Casimir operator commuting with all Black-Scholes LIVFs.  Since $X_{x}^{L}$ and $X_{p}^{L}$ do not commute, $C_{P}$ defines  two higher order 
polarizations.  The $x$-space polarization  is generated by $ \langle \, C_{P} ,  X_{p}^{L} \, \rangle$  
\begin{subequations}
\begin{align}
{X_{p}^L} \Psi  = 0 &  \rightarrow    \Psi = \zeta \, e^{ - p x }  \, \psi(x,t)  \label{hpolbs1} \\
X_C \Psi  = 0 &  \rightarrow   \nonumber \\
& \frac{\partial \psi}{\partial t} +  \frac{1}{2} \sigma^2 \frac{\partial^2  \psi}{\partial x^2}  + \mu \,  \frac{\partial  \psi}{\partial x}   - r \psi = 0  \label{hpolbs}
\end{align}
\end{subequations}
Equation \eqref{hpolbs} is the Black-Scholes equation in coordinate (price)  space.
\vskip 1mm
\subsection{Operators in Momentum Space}\label{FullQuant}
\vskip 1mm
\subsubsection{Right Invariant Vector Fields}
\vskip 1mm
\begin{align}
{X_{t}^R} & = \frac{\partial}{\partial t}  \qquad \qquad {X_{\zeta}^R}  \equiv  \Xi = \zeta \frac{\partial}{\partial \zeta}  \nonumber \\
{X_{x}^R}  & = \frac{\partial}{\partial x}    + p \Xi   \nonumber \\
{X_{p}^R} & =  \frac{\partial}{\partial p}  + \sigma^2 t  \frac{\partial}{\partial x}  + (\sigma^2 \,p  + \mu )\, t\, \Xi      
\label{rightXbs}
\end{align}
we have
\begin{align}
 [{X_{t}^R} , {X_p}^R] & = \sigma^2  {X_{x}^R} + \mu \, \Xi    &&[{X_{x}^R} , {X_p}^R] = \, \Xi 
\label{liebrakets01}
\end{align}
 All other brackets are zero{\footnote{
The inner product of the vertical form $\Theta$ and the RIVFs gives the (Galilean) Hamiltonian constants of motion
\begin{align}
\Theta ({X_{x}^R}) = 0 & \rightarrow p      \equiv p_0 \nonumber \\
\Theta ({X_{t}^R})  = 0  &     \rightarrow \frac{1}{2}\sigma^2 p^2+ \mu p  -r  \equiv E_0 \nonumber \\
\Theta ({X_{p}^R}) =0 &  \rightarrow x - ( \sigma^2  p + \mu  ) t    \equiv x_0  \nonumber
\label{noether}
\end{align}
}}. 
\vskip 1mm
Irreducibility is achieved
 by considering the action of the right invariant fields  on the {\it{constants of motion}}   $\Phi(p)$
in \eqref{pol1bs0}.
After straightforward algebra, one finds
\begin{equation}
  {X_{p}^R} : \Phi(p) \rightarrow   \frac{\partial}{\partial  p} \,\Phi(p) \quad 
  {X_{x}^R} : \Phi(p)  \rightarrow  \, p  \, \Phi(p)
\label{pol3bs2a}
\end{equation}
and
\begin{equation}
 - {X_{t}^R} : \Phi(p) \rightarrow ( \frac{1}{2 } \sigma^2 p^2  + \mu p  \,  -  r ) \, \Phi(p) 
\label{pol3bs29}
\end{equation}
From \eqref{pol3bs2a} the price operator, $\hat{x}$,  the {\it{momentum}} operator
$\hat{p}$  and the time evolution operator (Hamiltonian) $\hat{H}$ are given by
\begin{equation}
\hat{x} =  \frac{\partial}{\partial  p}  \quad \hat{p} =  p  \quad     \hat{H}  \equiv  \frac{1}{2} \sigma^2 {\hat{p}^2} + \mu   \hat{p}  - \, r 
\label{oper1}
\end{equation}
\vskip 1mm
\subsection{Operators in Coordinate Space}\label{OPX}
\vskip 1mm
The expressions \eqref{oper1} for the operators in momentum space
suggest the following definition for the 
coordinate operator, $\hat{x}$ and  the (non hermitian)  momentum , $\hat{p}$  operator in $x$-space
\begin{equation}
\hat{x} =  x  \quad \hat{p} =  \frac{\partial}{\partial x}  \qquad  [\hat{p} , \hat{x} ]=  1
\label{oper2}
\end{equation}
Equation \eqref{oper2} is a direct consequence of our extension of  the Heisenberg-Weyl subgroup by  $\mathbb{R}^{+}$ (equation \eqref{weyl3}).
The 2-dimensional translation subgroup  of the $WSp(2,  \mathbb{R} )$ group  acts on real exponentials, thus justifying the
definitions \eqref{oper2} and the  use for the inverse bilateral Laplace transform in  \eqref{pol1bs0}. 
This is an opposition to  quantum mechanics, where 
 the Heisenberg-Weyl subgroup acts on the unit circle instead, making the Fourier transform
the mapping between the coordinate and momentum spaces.
\vskip 1mm
Using equations \eqref{oper2}, the Black-Scholes Hamiltonian{\footnote{The non hermiticity of the Black-Scholes Hamiltonian  
has been extensively 
studied   in the literature (\cite{jana}, \cite{boson}, \cite{blasi}).
Non hermiticity  is relatively mild, with eigenvalues either real or appearing in complex conjugate pairs. }}
 in coordinate space is
\begin{equation}
{\hat{H}}_{BS}  \equiv  \frac{1}{2} \sigma^2 \frac{\partial^2}{\partial x^2} + \mu   \frac{\partial}{\partial x}  - \, r 
\label{oper3}
\end{equation}
\vskip 1mm
\subsubsection{Numeraire Coboundary Value}\label{MART}
\vskip 1mm
The value of the numeraire parameter $\mu$ can be determined by requiring the stock price $S \equiv e^x$  to be
a zero eigenvalue{\footnote{This is actually a martingale condition \cite{baaquie1}.}}
 of the Hamiltonian operator{\footnote{In momentum space, the zero eigenvalue condition reads
\begin{equation}
 (\frac{1}{2 } \sigma^2 p^2  + \mu p  - \,r) \Psi_{0} (p)   = 0 \rightarrow  \Psi_{0}(p) = \delta(\frac{1}{2 } \sigma^2 p^2  + \mu p - \,r)
\end{equation}
 It is straightforward to prove, using the properties of the Dirac delta and the inverse Laplace transform, that if we identify $\Psi_{0}(x) $ with the stock price $S$,
 $\Psi_{0}(x) \simeq \exp(x )$, this  requires  $ \mu  = r -\frac{1}{2} \sigma^2$
}}. 
\vskip 1mm
Using the  Black-Scholes Hamiltonian \eqref{oper3}, we find 
\begin{equation}
\qquad  {\hat{H}}_{BS} \, e^x = 0 \quad \Rightarrow \quad \mu  = r -\frac{1}{2} \sigma^2
\end{equation}
\vskip 1mm
\subsection{Pricing Kernel} \label{EXPECT}
\vskip 1mm
The kernel $K_{BS}(x,x^\prime, \tau)$ for the  Black-Scholes  equation \eqref{hpolbs}
\begin{equation*}
\Psi(\tau, x)   = \int K_{BS}(x,x^\prime, \tau) \Psi(0, x^\prime) d x
\end{equation*}
is obtained by an in inverse Laplace transform 
from the momentum representation in \eqref{momkernel}
\begin{equation}
 K_{BS}( x, x^\prime, t) =      \frac{1 }{2 \pi i} \, \int _{c-i \infty}^{c+ \infty } \, d p \, e^{  - E_{r}(p) \, t}  e^{p (x-x^\prime)}
 \label{pkernel1}
\end{equation}
The integral \eqref{pkernel1} does not exist for $t > 0$, in accordance to the fact  that pricing problems in finance  are time irreversible final value problems.
 \vskip 1mm
 Let $t < 0$ and define $\tau \equiv -t $ .Then the integral \eqref{pkernel1} exists and we recover the well known  Gaussian pricing kernel
\begin{equation}
\label{kernellag1}
 K_{BS}(x,x^\prime, \tau) =  e^{-r \tau} \,  \, \frac{1}{\sqrt{2 \pi \sigma^2 \tau}} e^{ -\frac{1}{2 \sigma^2 \tau } \, (x^\prime-x -\mu \tau)^2}
\end{equation}
\vskip 1mm
\subsubsection{Derivation using LCTs}
\vskip 1mm
The Black-Scholes pricing kernel can also be obtained with the methods developed in references \cite{Miller2} and \cite{wolf2}, using the properties of the  $WSp(2,  \mathbb{R} )$ group 
and its representation as Linear Canonical Transformations.
\vskip 1mm
From equation  \eqref{ltcdef2},
the $SL(2, \mathbb{R})$  matrix  \eqref{genmatrixBSa} generates  the LCT 
\begin{equation}
 W({\bf{M_{BS}}},x,x^\prime)  = \frac{1}{\sqrt{2 \pi \sigma^2 \,t }}\, \exp(  - ( x^\prime -x )^2/ (2 \,\sigma^2 \, t))
\end{equation}
which is the heat equation kernel (or Weierstrass transform $W[]$).
\vskip 1mm
$K_{BS}(x,x^\prime, \tau)$   is obtained by multiplying the heat  kernel $ W({\bf{M_{BS}}})$ by the discount factor $\exp (-r \tau)$ and setting $\tau = T-t$, $x \rightarrow  x + \mu \tau$.
We note that $x \rightarrow  x + \mu \tau$ is the Galilean transformation \eqref{bs1m2b} representing a numeraire change.
\vskip 1mm
 $K_{BS}$  can also be written in terms of the pseudo-differential operators {\footnote{
For smooth functions 
\begin{equation}
e^{{a\,{\frac{\partial}{\partial x}}}}f(x)=f(x+a)
\end{equation}
Combining the result above with the Gaussian integral
\begin{equation}
 e^{ a^2}  = {\frac  {1}{{\sqrt  {4\pi }}}}\int _{{-\infty }}^{\infty }  e^{a \, y} \,  e^{{-y^2}/4} dy 
\end{equation}
we get
\begin{equation}
 e^{   {\frac{\partial^2}{\partial x^2}}       } f(x) = {\frac  {1}{{\sqrt  {4\pi }}}}\int _{{-\infty }}^{\infty }  f(x-y) e^{-y^2} dy 
\end{equation}
}}
associated with $ W({\bf{M_{BS}}})$ \cite{wolf2}
\begin{align}
\label{kernellag2}
  e^{\tau \, {\hat{H}}_{BS}} \, f(x) & = \quad e^{- \, r \tau} \,  e^{ \frac{1}{2} \sigma^2 \tau  \frac{\partial^2}{\partial x^2}} \, e^{\mu  \tau  \frac{\partial}{\partial x}  } \, f(x) = \nonumber \\
&    \int _{- \infty}^{+ \infty }  d x^\prime  K_{BS} (x,x^\prime, \tau)  f(x^\prime) 
\end{align}

%% file: linear_interest_rates.tex

\section{Linear Potential. Ho-Lee Model} \label{IRQUANT}

\subsection{Quantization Group}
\vskip 1mm
The  $SL(2, \mathbb{R})$ parabolic shear mapping  \eqref{genmatrixBSa} that has been  used in the Black-Scholes  theory will be used in this section
for constructing the linear potential model.  
\vskip 1mm
The composition law  for the new quantization group   $\tilde{I}$ is 
\begin{align}
t^{\prime \prime} &  =t + t^\prime               \nonumber \\
 p^{\prime \prime} &  = p + p^{\prime }      \nonumber \\
 x^{\prime \prime} & =x + x^{\prime }  +  \sigma^2 p^{\prime} t       \nonumber \\
\zeta^{\prime \prime} & =\zeta^{\prime } \zeta e^{   \epsilon_{IR}(g,g^\prime) }   
\label{cocycleIRa}
\end{align}
where $ t,p,x\in {\mathbb{R}}$ and $\zeta \in {\mathbb{R}}^{+}$ 
with the cocycle  $\epsilon_{IR}$ 
\begin{align}
\epsilon_{IR}(g,g^{\prime})  & =  \epsilon_{G}(g,g^{\prime})  \, +    \,  \epsilon_{N}(g,g^{\prime})  + \,  \epsilon_{I}(g,g^{\prime}) 
\end{align}
$\epsilon_{G}(g,g^{\prime})$ is the Galilean cocycle in equation \eqref{galileancocycle} and  $\epsilon_{N}(g,g^{\prime}) $ is the coboundary term given in \eqref{galileancob}. 
We have added a new true cocycle  (not a coboundary) $\epsilon_{\beta}(g,g^{\prime}) $ 
\begin{equation}
\epsilon_{\beta}(g,g^{\prime})  =  \beta\, t \,  (x^{\prime}   + \frac{1}{2} \sigma^2  {p^\prime}  t ) \qquad  \qquad \beta \in {\mathbb{R}}
 \label{cocyclebs2ira}
\end{equation}.
\vskip 1mm
We interpret the group coordinates{\footnote{Identifying group parameters is usually done after analyzing the polarization, however in this case the identification is simple enough.}} as $x$ representing a short rate,
$p$ its conjugate momentum, and $t$ the calendar time.
The parameter $\sigma$ is the short rate volatility.
\vskip 1mm
We will verify in the next sections that this group describes the Ho–Lee interest rate model. 
In quantum physics, the $U(1)$ extension of this group describes the free fall (linear potential).
The Group Quantization formalism applied to the linear potential in physics can be found in \cite{wolf11}.
\vskip 1mm
\subsection{Lie Algrebra}{\label{IRVF}}
\vskip 1mm
\subsubsection{Left Invariant Vector Fields}
\vskip 1mm
From the composition  law  \eqref{cocycleIRa}  and  \eqref{cocyclebs2ira}, we obtain
\begin{align}
{X_{\zeta}^L}  & \equiv  \Xi = \zeta \frac{\partial}{\partial \zeta}  \nonumber \\
 {X_{x}^L}  & = \frac{\partial}{\partial x}   \qquad  \qquad  {X_{p}^L}  = \frac{\partial}{\partial p} + x \, \Xi  \nonumber \\
{X_{t}^L}  & = \frac{\partial}{\partial t} + \sigma^2  p  \frac{\partial}{\partial x}  +(E(p)+ \beta x )\, \Xi  
\label{leftXir}
\end{align}
where
\begin{equation}
E (p)  \equiv  \frac{1}{2} \, \sigma^2  p^2  + \mu \, p
\end{equation}
with the non-zero Lie Brackets
\begin{align}
[{X_{t}^L} , {X_{x}^L} ]  & = -\beta \, \Xi   & [{X_{t}^L} , {X_{p}^L} ]  &  = - \sigma^2   {X_{x}^L}   -\mu \Xi \nonumber \\
[{X_{p}^L} , {X_{x}^L} ]  & = - \Xi & \qquad
\label{liebraketsir0}
\end{align}
Note that, as opposed  to the Black-Scholes case (section \ref{IVF}), the time generator and the $x$-generator do not commute.
\vskip 1mm
\subsection{Connection}
\vskip 1mm
The vertical form $\Theta$ and the curvature form $ d \Theta $  are  given by
\begin{align}
& \Theta  =      -x \,d p  -   E(p) \, dt  -\beta \,x \,dt + d \Xi   \nonumber \\
& d \Theta =   d p \wedge  d x   - ( \sigma^2  p  + \mu) \, dp \wedge dt     -\beta \, d x \wedge  d t 
\label{theta2ir}
\end{align}
with 
\begin{equation}
 d\, \Xi  \equiv\frac{1}{ \zeta} d \zeta  
\end{equation}
\vskip 1mm
\subsubsection{Characteristic Module}\label{charmod}
\vskip 1mm
$\mathcal{C}_\Theta$ 
is spanned by a unique field $X_C $
\begin{align}
& X_C =   {X_{t}^L}  + \mu \, {X_{q}^L}  - \beta \, {X_{p}^L} = \\ \nonumber
 &     \frac{\partial}{\partial t}   - \beta   \frac{\partial}{\partial p}  +  (p\,\sigma^2  + \mu) \, \frac{\partial}{\partial x}  +  ( \frac{1}{2} \,\sigma^2 \, p^2 + \mu p)  \, \Xi 
\label{mod2bbir}
\end{align}
\vskip 1mm
\subsection{First Order Polarization}{\label{FPOIR}
\vskip 1mm
\subsubsection{Ho-Lee Equation in Momentum Space}
\vskip 1mm
As in the Black-Scholes case, there is only one  first order $p$-space polarization $\mathcal{P}$ spanned 
by the $x$-translations  and the  $\mathcal{C}_\Theta$ generator $X_C$.   
\vskip 1mm
The polarized functions $\Psi \in {\mathcal{F_P}}$ are found by imposing the polarization constraints
on the functional space ${\mathcal{F_P}}$
\begin{equation*}
{\mathcal{F_P}} = \{\, \Psi : \tilde{I} \rightarrow \mathbb{R} \, / \,  \Psi(x,p, t,\zeta) = \zeta \Psi(p,q,t)  \, \}
\end{equation*}
\vskip 1mm
One has  
\begin{subequations}
\begin{align}
{X_{x}^L} \Psi  = 0 &  \rightarrow    \Psi = \zeta \, \psi(p,t)   \\
X_C \Psi  = 0 &  \rightarrow   \nonumber \\
& \frac{\partial \psi}{\partial t} - \beta \, \frac{\partial \psi}{\partial p} + \frac{1}{2} \sigma^2 p^2 \psi +  \mu \, p \, \psi = 0 \label{firstpolir}
\end{align}
\end{subequations}
Equation \eqref{firstpolir} is the Ho-Lee Equation in Momentum Space.
\vskip 1mm
The general solution{\footnote{We will show in section \ref{HOP2} that the polarized functions can be expressed in the $x$-space using Airy functions. This can be anticipated by the identity \cite{abra}
\begin{equation*}
 \frac{1 }{2 \pi i} \, \int _{c-i \infty}^{c+ \infty } \, d p \,   e^{x p + t p^3} = \frac{1}{  (3 t)^{\frac{1}{3}}} \, Ai\left( \frac{-x}{ (3 t)^{\frac{1}{3}}} \right) \qquad c>0
\end{equation*}
The solutions that are not proportional to $Ai$ grow very rapidly at infinity and they need to be constructed
with different contours  in the complex plane.}} 
of  \eqref{firstpolir} is
\begin{equation}
\psi(p, t) = \Phi(p+\beta t) \, e^{ \frac{1}{2 \beta} \mu p^2 +  \frac{1}{6 \beta} \sigma^2 p^3}
\label{firstpolir0}
\end{equation}
where $\Phi(p)$ is an arbitrary function of the momentum $p$.
\vskip 1mm
The bilateral Laplace transform relates the $p$-space and the $x$-space. From equation \eqref{firstpolir0}, the  general expression of a polarized function in $x$-space can be written as
\begin{equation}
 \Psi(\zeta, x, t) =   \zeta \, e^{ - \beta t x} \,   \frac{1 }{2 \pi i} \, \int _{c-i \infty}^{c+ \infty } \, dp\,   G(p ,t) \,  e^{ x p} \, \Phi(p) 
\label{price1bb}
\end{equation}
where  
\begin{equation}
 G (p, ,t) \equiv e^{\frac{1}{2 \beta } \mu (p-\beta t)^2 +  \frac{1}{6 \beta} \sigma^2 (p-\beta t)^3}
\label{price1bb2}
\end{equation}
The integration contour is   a
vertical line on the complex plane, such that all singularities of the integrand lie on the left of it.
\vskip 1mm
\subsection{High Order Polarization.}\label{HOP2} \label{HOLEEEQ}
\vskip 1mm
\subsubsection{Ho-Lee Equation in Coordinate Space}
\vskip 1mm
The pricing (evolution) equation can be found directly in $x$-space by using a  {\it{high-order}} polarization (see section {\ref{SecondPO}}).
\vskip 1mm
Let   $X_C$  be the  $\mathcal{C}_\Theta$  generator. The second order operator 
\begin{equation}
X_{P}  = X_C +  \frac{1}{2}\,\sigma^2 \,  X_{x}^{L} \,X_{x}^{L}  
\label{cas1ir}
\end{equation}
is a Casimir operator commuting with all LIVFs (equation \refeq{leftXir}). Since $X_{x}^{L}$ and $X_{p}^{L}$ do not commute, $X_{P}^{L}$ defines  two higher order 
polarizations, the  $p$-space polarization ${\mathcal{P}}_{p} = \{ \, X_{P}^{L} ,  X_{x}^{L} \}$, and the $x$-space  polarization ${\mathcal{P}}_{x}  = \{ \, X_{P}^{L} ,  X_{p}^{L} \}$.
\vskip 1mm
The $x$-space polarized functions are obtained by imposing the  ${\mathcal{P}}_{x}$ polarization constraints on functions (sections)  of the form $\Psi(z,p,x,t) = \zeta \Psi(p,x,t)$
\begin{subequations}
\begin{align}
{X_{p}^L} \Psi  = 0 &  \rightarrow    \Psi = \zeta  \, V (x,t)    \\
X_P \Psi  = 0 &  \rightarrow   \nonumber  \\
&  \frac{\partial V}{\partial t} + \frac{1}{2} \sigma^2 \frac{\partial^2 V}{\partial x^2} +  \mu \frac{\partial V}{\partial x}+  \beta \, x \, V = 0  \label{highpolir}
\end{align}
\end{subequations}
Equation \eqref{highpolir} is the Ho-Lee equation in coordinate space.
\vskip 1mm
If we write
\begin{equation}
V(x,t) = \,  \frac{1 }{2 \pi i} \, \int _{c-i \infty}^{c+ \infty } \, dp\,  \Phi(p,t) \, e ^{p x}   
\label{highpolir2}
\end{equation}
and substitute in \eqref{highpolir}, we recover  the first order polarization equations \eqref{firstpolir}.
\vskip 1mm
Using the Feyman-Kac{\footnote{
Consider the differential operator $L$ 
\begin{equation*}
L \equiv \frac{1}{2} {\omega^2}(x,t)\frac{\partial^2}{\partial x^2} +  \mu(x,t)\frac{\partial}{\partial x}
\end{equation*}
where $\mu(x, t), \omega(x, t)$  and $r(x, t)$ are functions of $(x, t)$ , with $x$ defined in a real 
domain $D \subset  \mathbb{R}$ and $t$ a positive real number, $t \in   [0, T] $
Then, subject to technical conditions, the unique solution of the PDE
\begin{equation*}
(\frac{\partial}{\partial t} + L  + r(x,t) ) \, f(x,t) = 0 \quad x \in D, 0 \geq  t  \leq  T
\end{equation*}
with terminal value $f(x, T ) = g(x)$, is given by  the {\it{Feynman-Kac Formula}}
\begin{equation*}
f(x,t) = \mathbb{E} ( e^{- \int_t^{T} r(X,s) ds }\left.  g(X_T) \right|  X(t) = x)
\end{equation*}
where the expectation is taken with respect to the transition density induced
by the SDE
\begin{equation*}
 d X = \mu(X, t) dt + \omega (X, t) d W
 \end{equation*}
 with $W$ a Brownian motion.
}} theorem, the  the solution of  \eqref{highpolir} can be written as
\begin{equation}
V(x,t) = \mathbb{E} ( e^{- \beta \int_t^{T} X(s) ds }\left.  \phi(X_T) \right|  X(t) = x)
\label{fkac}
\end{equation}
where the expectation is taken with respect to a normal process with volatility $\sigma$ and 
drift $\mu$   ($W$ is a Brownian montion)
\begin{equation*}
 d X = \mu dt + \sigma\, d W
 \end{equation*}
\vskip 1mm
\subsection{Airy Expansion}
\vskip 1mm
After performing the following change of function and independent variable 
\begin{align}
& V(x,t) \equiv   e^{-\frac{\mu}{\sigma^2}\, x} \,  e^{\lambda  \,t} U(y) \\ \nonumber
& y \equiv \left(\frac{2}{\sigma^2} \right)^{\frac{1}{3}} \, ( \frac{\mu}{2 \sigma^2} - \lambda - \beta \, x )
\label{airy4}
\end{align}
the polarization equation \eqref{highpolir} becomes the Airy equation 
\begin{equation}
\frac{\partial^2 U}{\partial y^2} - y \, U(y) = 0
\label{airy3}
\end{equation}
Therefore,  the polarized functions can then be written as a superposition of Airy functions{\footnote{There is a distinguished solution of equation \eqref{airy3},
called $Ai$, that decays rapidly as $y \to + \infty$, while a second linearly independent solution $Bi$ grows rapidly in this limit. Also, like Bessel functions,
both  $Ai$ and  $Bi$ are oscillatory with a slow decay for large values of their arguments. See reference \cite{abra} }} 
\begin{equation}
V(x,t) =    e^{-\frac{\mu}{\sigma^2}\, x} \, \sum_{i=0}^{\infty} \, e^{\lambda_i \,t} \, (  a_i \, Ai(y_i)  +   b_i \,  Bi(y_i))   
\label{highpolir4}
\end{equation}
where $  a_i, b_i, \lambda_i \in {\mathbb{R}}$ and
\begin{equation*}
y_i \equiv (\frac{2}{\sigma^2})^{\frac{1}{3}} \, (\frac{\mu}{2 \sigma^2} - \lambda_i - \beta \, x)
\end{equation*}
\vskip 1mm
This form is convenient for the analysis of complex boundary conditions in bond pricing problems, as shown in \cite{goldstein}. The boundary conditions constrain the values 
for the expansion 
coefficients $a_i, b_i$ and $\lambda_i$.
\vskip 1mm
Note that 
the group contraction $\beta \rightarrow 0$ is smooth and the group law \eqref{cocycleIRa} reduces to the Black-Scholes group law,
however $\beta = 0$ is a singular point in the integral for the momentum polarized functions,   \eqref{firstpolir0}
and the polarization equation  \eqref{firstpolir}. 

\vskip 1mm
\subsection{Pricing Kernel}
\vskip 1mm
\subsubsection{Similarity Methods}
\vskip 1mm
One can use the methods developed in  \cite{Miller2} and \cite{wolf2} in order to find coordinate changes
mapping  Black-Scholes solutions into solutions of the  equation  \eqref{highpolir}
\vskip 1mm
For instance, it can be shown that if
$V_{BS}(x,t)$ a solution of the Black-Scholes equation \eqref{hpolbs} with $r = 0$,  then 
\begin{equation}
V(x,t) =   \Omega_\beta(x,t) \,  \, V_{BS}( x - \frac{1}{2} \beta \sigma^2 t^2 ,t) \
\end{equation}
with
\begin{equation}
\Omega_\beta(x,t) = e^{\beta x t + \frac{1}{6} \beta^2 \sigma^2 t^3  + \frac{1}{2} \mu \beta t^2 }
\label{omega1}
\end{equation}
is a solution of \eqref{highpolir}. Hence, the Ho-Lee pricing Kernel can be obtained from the Black-Scholes pricing Kernel \eqref{kernellag1} (with $r = 0$)  in an analogous manner
\begin{equation}
\label{kernellag2}
 K_I(x,x^\prime, \tau) =   \, \Omega_\beta(x,\tau) \, K_{BS}(x -  \frac{1}{2} \beta \sigma^2 \tau^2, x^\prime, \tau)
\end{equation}
where we have switched to the variable $\tau = T-t$.
\vskip 1mm
Notice that $\Omega_\beta(x,\tau)$ is a solution of \eqref{highpolir}. In fact, $\Omega_1(x,\tau) $  is the expression for the Ho-Lee bond maturing at $T$
\begin{equation}
\Omega_1(x,\tau) = e^{-x (T-t) - \frac{\sigma^2}{6} (T-t)^3 + \frac{\mu}{2} (T-t)^2}  \qquad  t \in [0, T]
\end{equation}
\vskip 1mm
\subsubsection{Pseudo-Differential Operators}\label{EXPECT2}
\vskip 1mm
 The pseudo-differential operator methods in section \ref{EXPECT} can also  be used 
 to compute the evolution operator $U$. 
 \vskip 1mm
 From equation \eqref{highpolir}.
\begin{equation}
U(\tau) =  \exp(\tau  H_I)    \qquad   H_I \equiv \frac{1}{2} \sigma^2 \frac{\partial^2 }{\partial x^2} +  \mu \frac{\partial }{\partial x} +  \beta \, x 
\end{equation}
where $\tau = -t$.  We split $H_I$ into two operators
\begin{equation}
 A  \equiv \frac{1}{2} \sigma^2 \frac{\partial^2 }{\partial x^2}  + \mu \frac{\partial }{\partial x} \quad \qquad   B =    \beta \, x 
\end{equation}
Given that the only non zero commutators are
\begin{equation*}
 [A,B]  = \beta \sigma^2 \frac{\partial }{\partial x} + \mu\,\beta \qquad [B,[A,B]]  = -\beta^2 \, \sigma^2
\end{equation*}
it is possible to apply the {\it{left oriented extended}} version of the
Baker-Campbell-Hausdorff formula  \cite{zazen}
\begin{equation}
 e^{\tau A + \tau B} =  e^{\frac{1}{3!} \tau^3 ( 2 [B,[A,B]] + [A,[A,B]])} \, e^{\frac{1}{2} \tau^2 [A,B]}  \,  e^{\tau B} \, e^{\tau A}
\end{equation}
We obtain{\footnote{We have interchanged the exponential factors
\begin{equation*}
e^{\frac{1}{2} \tau^2  \beta \sigma^2 \frac{\partial }{\partial x}} \,  e^{\beta \tau x} 
\end{equation*}
since the commutator is a number
\begin{equation}
 e^{\frac{1}{2} \tau^2  \beta \sigma^2 \frac{\partial }{\partial x}} \,  e^{\beta \tau x} = e^{\frac{1}{2} \tau^3  \beta^2 \sigma^2 }\, e^{\beta \tau x} \, e^{\frac{1}{2} \tau^2  \beta \sigma^2 \frac{\partial }{\partial x}} 
\end{equation}}}
\begin{align}
e^{\tau H_I} & =    e^{- \frac{1}{3} \beta^2 \, \sigma^2  \tau^3} \, e^{\frac{1}{2} \mu\,\beta \tau^2}  \nonumber \\
&  \times \,\, \, e^{\frac{1}{2} \tau^2  \beta \sigma^2 \frac{\partial }{\partial x}}  \,  e^{\beta \tau x} \, 
e^{\frac{1}{2} \sigma^2 \tau \, \frac{\partial^2 }{\partial x^2} } \, e^{\mu \tau \, \frac{\partial }{\partial x} }  \nonumber \\
&  =  \Omega_{\beta}(x,\tau) \,  e^{\frac{1}{2} \tau^2  \beta \sigma^2 \frac{\partial }{\partial x}}  \, e^{\frac{1}{2} \sigma^2 \tau \, \frac{\partial^2 }{\partial x^2} }\, e^{\mu \tau \, \frac{\partial }{\partial x} }
\label{operlo}
\end{align}
where $\Omega_{\beta}(x,\tau) $ has been defined in \eqref{omega1}. 
We apply the operators in \eqref{operlo} from right to left, and use  the Weierstrass theorem to recover equation \eqref{kernellag2}
\begin{equation}
\label{kernellag2a}
 K_I(x,x^\prime, \tau) =   \Omega(x,\tau)  \, K_{BS}(x -  \frac{1}{2} \beta \sigma^2 \tau^2, x^\prime, \tau)
\end{equation}

%% file: harmonic_oscillator.tex


As described in section \ref{HWGROUP},
our version of the  Heisenberg-Weyl group consists in the 2-dimensional translations centrally extended by $\mathbb{R}^{+}$, such that
the commutator between the momentum
and  coordinate operators is $[\hat{p},\hat{x}]= 1$.  
\vskip 1mm
We will verify in this section that in our modified version  of the  $WSp(2,  \mathbb{R} )$ group,  the hyperbolic  $SL(2, \mathbb{R})$ subgroup generates the harmonic oscillator 
quantization group $\tilde{H}$ 
 \begin{equation}
 {\bf{M_H}} =  \begin{pmatrix*}[c]
    \phantom{\lambda^{2}}  \cosh \Omeg t &  \, \lambda^{-2}  \sinh \Omeg t\\
    \lambda^{2} \sinh \Omeg t  &  \phantom{ \lambda^{-2}  } \cosh \Omeg t
\end{pmatrix*} 
\qquad \lambda \equiv \frac{\sqrt{\Omeg}} {\Sig} 
\label{genmatrixHO}
\end{equation}
where $\Omeg, \Sig, t \in \mathbb{R}$
\vskip 1mm
By contrast, in quantum mechanics  (\cite{garcia0}, \cite{bisquert}),
where  the 2-dimensional translations are centrally extended by $U(1)$, the commutator between the momentum
and  coordinate operators is pure imaginary, $[\hat{p},\hat{x}]= -i \, \hslash$, and the 
harmonic oscillator quantization group is generated by an elliptic $SL(2, \mathbb{R})$ subgroup
 \begin{equation}
 \begin{pmatrix*}[c]
    \phantom{-\lambda^{2}}  \cos \Omeg t &  \, \lambda^{-2}  \sin \Omeg t\\
    -\lambda^{2} \sin \Omeg t  &  \phantom{ \lambda^{-2}  } \cos\Omeg t
\end{pmatrix*}
\label{genmatrixHOa}
 \end{equation}
The  $SL(2, \mathbb{R})$  matrix \eqref{genmatrixHO} acts as a squeeze mapping of the Euclidean plane,
whereas  \eqref{genmatrixHOa}  corresponds to a rotation. Both subgroups can be transformed into each other by the
the mapping $\omega \rightarrow i \, \omega$. Notice that the mapping between subgroups is not the {\it{Euclidean}} time rotation   $t \rightarrow i \, t$.
\vskip 1mm
\subsubsection{Group Composition Law}
\vskip 1mm
Using equation  \eqref{slr2a},  ${\bf{M_H}}$  generates  the following composition law:
\begin{align}
 t^{\prime \prime} & =t + t^\prime  \nonumber \\
p^{\prime \prime}& = p + p^\prime \,  \cosh \Omeg t  +  \lambda^{2}   x^\prime \sinh \Omeg t  \nonumber \\
 x^{\prime \prime} &  = x +   x^{\prime}   \, \cosh \Omeg t  + \lambda^{-2}  \, p^\prime \,  \sinh \Omeg t     \nonumber \\
 \zeta^{\prime \prime} & = \zeta^\prime \zeta \exp \left (  \epsilon_H(g^\prime, g) \right)
\label{cocyclebs1z}
\end{align}
with the cocycle  $\epsilon_{H}$ 
\begin{align}
\epsilon_{H}(g,g^{\prime}) & = \frac{1}{2} ( p x^\prime - x p^\prime) \cosh \Omeg t+ \\ \nonumber
 & \quad \frac{1}{2} (\lambda^{-2} p \, p^\prime - \lambda^{2} \, x \,  x^\prime)  \sinh \Omeg t 
 \label{cocyclehs2}
\end{align}
The matrix ${\bf{M_H}}$  reduces to the Black-Scholes generating matrix   ${\bf{M_{BS}}}$    in the limit $w \rightarrow 0$, and correspondingly,
the quantization group $\tilde{H}$ contracts to the Black-Scholes quantization group $\tilde{G}$ (modulo coboundaries)  in this limit.
For brevity, we have omitted the  numeraire coboundary{\footnote{The generating function for this coboundary is $\mu \, x$, and the final expression is  
$\mu ( \lambda^{-2} p^\prime \sinh  \Omeg t  + x^{\prime} \cosh \Omeg t-  x^{\prime})$, with $\mu$ real.}}  \eqref{galileancob} in the composition law \eqref{cocyclebs1z}.
\vskip1mm
\subsection{Polarization using Orthogonal Coordinates}\label{LADDERPOL}
\vskip1mm
The  $SL(2, \mathbb{R})$ matrix ${\bf{R}}$  
 \begin{equation}
\qquad \quad {\bf{R}} = \frac{1}{\sqrt{2}} \, \begin{pmatrix*}[c]
    \lambda  &  - 1/ \lambda  \\
   \lambda  &    \phantom{-} 1/ \lambda 
\end{pmatrix*}  
\label{diagonalM}
\end{equation}
transforms  ${\bf{M_H}}$ into a change of scale (squeeze) operator
 \begin{equation*}
 {\bf{D_H}} = {\bf{R^{-1}}}  \, {\bf{M_H}} \, {\bf{R}} =  \begin{bmatrix}
    e^{  \Omeg t} &  0  \\
    0 &     e^{ -  \Omeg t} 
\end{bmatrix}  
\end{equation*}
The change of coordinates 
\begin{align}
A =& \frac{1}{\sqrt{2}} (  \frac{1}{\lambda} \, p -  \lambda \,  x)  \nonumber \\
B =& \frac{1}{\sqrt{2}} (   \frac{1}{\lambda} \, p  + \lambda \, x )
\label{diagonalFrac}
\end{align}
splits  the phase space (p,x) into orthogonal subspaces. The group law \eqref{cocyclebs1z} is written as 
\begin{align*}
 t^{\prime \prime} & =t + t^\prime  \nonumber \\
 A^{\prime \prime} &  = A +   A^{\prime}   e^{-\Omeg t} \\
B^{\prime \prime} &  = B +   B^{\prime}   e^{\Omeg t}  \nonumber \\
 \zeta^{\prime \prime} & = \zeta^\prime \zeta \, e^{  \epsilon_H(g^\prime, g) }
\end{align*}
with
\begin{equation*}
\epsilon_{H}(g,g^{\prime})   =\frac{1}{2} (B^{\prime} A \, e^{ \Omeg t }\, -  \, B \, A^{\prime} \, e^{ - \Omeg t }) 
\end{equation*}
\vskip 1mm
\subsubsection*{Left Invariant Vector Fields}
\vskip 1mm
\begin{align*}
{X_{t}^L}  & = \frac{\partial}{\partial t} -\Omeg \, A \frac{\partial}{\partial A}  + \Omeg \, B \frac{\partial}{\partial B}   \nonumber \\
{X_{A}^L} & = \frac{\partial}{\partial A}   + \frac{1}{2} B \, \Xi  \qquad {X_{B}^L}  = \frac{\partial}{\partial B}  - \frac{1}{2} A \, \Xi
\end{align*}
with Lie Brackets
\begin{align*}
[{X_{t}^L} , {X_{A}^L} ]  & = w   {X_{A}^L}   \qquad  \quad [{X_{t}^L} , {X_{B}^L} ]    = - w   {X_{B}^L}   \nonumber \\
[{X_{A}^L} , {X_{B}^L} ]  & = -\Xi
\end{align*}
\vskip 1mm
\subsection*{Connection}
\vskip 1mm
The vertical form $\Theta$  and the curvature form $ d \Theta $  are  given by
\begin{align}
\Theta  & =   \frac{1}{2} (A \,d B  - B \, d A ) -   w A B  \, d t + d \Xi \nonumber \\
d \,\Theta &  =   d A \wedge  d B  -   w A \, d B  \wedge d t  -    w B \, d A  \wedge d t
\end{align}
\vskip 1mm
$\mathcal{C}_\Theta$ is spanned by the time generator  ${X_{t}^L} $.
\vskip 1mm
\subsection*{First Order Polarization}
\vskip 1mm
We choose the first order polarization algebra spanned by
\begin{equation*}
{\mathcal{P}} = \langle {X_{B}^L} , {X_{t}^L}     \rangle
\end{equation*}
The polarized functions $\Psi \equiv \zeta \, \Psi(A,B,t) $ are found by imposing 
\begin{subequations}
\begin{align}
 {X_{B}^L} \, \Psi & = 0 \quad   \rightarrow \quad  \Psi = \zeta \, e^{\frac{A B}{2}} \Psi(A,t) \nonumber  \\
{X_{t}^L} \, \Psi & = 0 \quad   \rightarrow  \quad  \frac{\partial \Psi}{\partial t}   -   \Omeg A \, \frac{\partial \Psi}{\partial A}   = 0 \label{ladder1}
\end{align}
\end{subequations}
The readers familiar with quantum mechanics will find 
equation \eqref{ladder1} formally similar to quantum harmonic oscillator coherent state equations in Fock space  (\cite{garcia0}, \cite{bisquert}), except that here the coordinate $A$ is real, not complex.
\vskip 1mm
We give the solution  of these polarization constraints in equation \eqref{coherent2}.
Equation \eqref{ladder1} coincides with equation \eqref{polho1a} obtained in the next section \ref{PHASEPOL}. 
\vskip 1mm
\subsection{First Polarization in Phase Space }\label{PHASEPOL}
\vskip 1mm
We use now the original phase space coordinates $(p,q)$ in the composition law \eqref{cocyclebs1z}.
\vskip 1mm
Since the generating matrix $M$ does not leave invariant either the $p$ or the $x$ space,  a first order polarization  cannot map the phase space  $(p, x)$ into either the momentum or the coordinate space.
However, the results in this section will be illustrative of a phase space formulation for the harmonic oscillator and will also be useful for finding a higher order polarization.
\vskip 1mm
\subsubsection*{Left Invariant Vector Fields}
\vskip 1mm
\begin{align}
{X_{t}^L}  & = \frac{\partial}{\partial t} +  \Omeg \, \lambda^2 x  \, \frac{\partial}{\partial p}  + \Omeg \, \lambda^{-2} p   \, \frac{\partial}{\partial x}   \nonumber \\
{X_{x}^L} & = \frac{\partial}{\partial x} - \frac{1}{2} p \, \Xi  \nonumber \quad  \nonumber \\
{X_{p}^L} & = \frac{\partial}{\partial p}  + \frac{1}{2} x \, \Xi
\label{leftXHO1}
\end{align}
\begin{align}
[{X_{t}^L} , {X_{p}^L} ] &  = - \Omeg \, \lambda^{-2} {X_{x}^L}    \qquad   [{X_{t}^L} , {X_{x}^L} ]  =- \Omeg \, \lambda^2 \, {X_{p}^L}   \nonumber \\
[{X_{p}^L} , {X_{x}^L} ] & = \,-  \Xi 
\label{liebraketsHO}
\end{align}
 All other brackets are zero. 
 \vskip 1mm
\subsection*{Connection}
\vskip 1mm
\begin{align}
\Theta & =    \frac{1}{2}(p dx - x dp) - E(p,x) dt+   d \, \Xi    \nonumber \\
d \Theta & =   d p \wedge  d x   -   \Omeg \, \lambda^{-2} \,  p \, dp \wedge dt  +  \Omeg \, \lambda^2  \, x \, dx \wedge dt
\label{theta2bs}
\end{align}
with
\begin{align}
 & E(p,x) \equiv  \frac{1}{2 } \Omeg ( \lambda^{-2} p^2  - \lambda^2 \, x^2 ) \nonumber \\
& d\, \Xi  \equiv\frac{1}{ \zeta} d \zeta 
 \label{energyHO}
 \end{align}
\vskip 1mm
The time translations ${X_{t}^L}$ are the only generator of the characteristic module $\mathcal{C}_\Theta$.
\vskip 1mm
\subsection*{First Order Polarization}
\vskip 1mm
There are two first order polarizations $ \mathcal{P}$ in phase space
\begin{subequations}
\begin{align}
&\langle {X_{t}^L}, Y_{+} \rangle  \qquad  / \quad   Y_{+} \equiv \, \lambda \, {X_{p}^L} +  \lambda^{-1} \, {X_{x}^L}    \label{hopol1} \\
& \langle {X_{t}^L}, Y_{-}   \rangle  \qquad  / \quad  Y_{-} \equiv   \,\lambda \, {X_{p}^L} -  \lambda^{-1} {X_{x}^L}  
\end{align}
\end{subequations}
We choose \eqref{hopol1} as polarization. Let $ \Psi$ a function of the form $ \Psi(\zeta, p, q, t) = \zeta \, \Psi( p, q, t) $
\begin{align}
& Y_{+} \Psi  = 0   \rightarrow   \nonumber  \\
& \lambda  \frac{\partial \Psi}{\partial p} -   \lambda^{-1}  \frac{\partial \Psi}{\partial x} + \frac{1}{2}\,  (\lambda x -  \lambda^{-1} p) \Psi = 0    \label{phasecons5a}
\end{align}
The solution of   \eqref{phasecons5a} is
\begin{equation}
\Psi(x,p, t,\zeta)  = \zeta   \psi ( A, t )\, e^{ \frac{1}{2} A \, B}
 \label{phasecons5}
\end{equation}
where  $A = A(p,q)$ and $B=B(p,q)$ are the orthogonal coordinates  \eqref{diagonalFrac}. Then
\begin{equation}
 {X_{t}^L} \Psi  = 0   \rightarrow  
\frac{\partial \psi}{\partial t} -  \Omeg \, A \,  \frac{\partial \psi}{\partial A}  = 0 
\label{polho1a}
\end{equation}
The constraint \eqref{polho1a} is the previously found  \eqref{ladder1} using orthogonal coordinates.
\vskip 1mm
\subsection{First Order Polarized Functions}\label{phasepol}
\vskip 1mm
Using the separable solution of \eqref{polho1a}
\begin{equation*}
\qquad  \qquad   \,  e^{\Omeg \, n \, t} \, A^n e^{\Omeg n t} \qquad n  \in \mathbb{N}^{+}
\end{equation*}
 the polarized functions are expressed as an infinite series
\begin{equation}
\Psi(\zeta, p,x,t) = \zeta \, \sum_{n=0}^{\infty} e^{\Omeg \, n \, t} \,   \Alp_n   \, A^{n} e^{\frac{1}{2} A B}
\label{coherent2}
\end{equation}
where $\Alp_n \in \mathbb{R}$ are the expansion coefficients.
\vskip 1mm
First order polarization in non orthogonal $(p,q)$ coordinates leads naturally to a financial theory in phase space, with 
\begin{equation*}
 F(p, q) = e^{\frac{1}{2} A B} = e^{\frac{1}{4} (\lambda^{-2} \, p^2 - \lambda^2 \, x^2)}
\end{equation*}
the analog of a Husimi quasi-probability. 
The expression for the  quantum harmonic  coherent states in the Fock basis and the Bargmann-Segal transform
can be obtained{\footnote{( The zero energy $1/2$ is the change or coordinates in the fiber, $\zeta\rightarrow \zeta \exp( t/2)$}}
from  \eqref{coherent2} by mapping the phase space $(p,q)$ into $\mathbb{C}$ by the simple correspondence  $p \rightarrow i p$.
\vskip 1mm

\subsection{High Order Polarization}\label{HOP}
\vskip 1mm
\subsubsection{Harmonic Oscillator Equation in Coordinate Space}
\vskip 1mm
The pricing equation  can be obtained directly in $x$-space by using a high-order polarization.
\vskip 1mm
The second order operator 
\begin{equation*}
X_{P}  = X_ t +  \frac{1}{2} \Omeg \lambda^{-2} \,  X_{x}^{L} \,X_{x}^{L}   - \frac{1}{2} \Omeg \lambda^{2} \,  X_{p}^{L} \,X_{p}^{L} 
\label{cas1ho}
\end{equation*}
is a Casimir operator commuting with all LIVFs.  
\vskip 1mm
Since $X_{x}^{L}$ and $X_{p}^{L}$ do not commute, $X_{P}$ defines  two higher order 
polarizations, $\langle X_{P} ,  X_{x}^{L} \rangle$ and $\langle X_{P} ,  X_{p}^{L} \rangle$.  
The coordinate space  representation is generated by $\langle X_{P} ,  X_{p}^{L} \rangle$
\begin{subequations}
\label{highpolhoALL}
\begin{align}
X_{p}^{L}  \Psi & = 0   \rightarrow \qquad  \Psi (\zeta,p,q,t) = \zeta e^{- p x/2} \, \psi(x,t)  \nonumber \\
 X_{P}  \Psi & = 0    \rightarrow   \nonumber \\
&   \frac{\partial \psi}{\partial t}  + \frac{1}{2} \Omeg \left( \lambda^{-2} \frac{\partial^2  \psi}{\partial x^2}   -   \lambda^2 \, x^2 \, \psi  \right)= 0 \label{highpolho}
\end{align}
\end{subequations}
\vskip 1mm
Since in finance one is usually interested in final value problems, we make the change 
$\tau = T - t$, where $T$ is a {\it{maturity time}} and $t \le T$.  Equation \eqref{highpolho} becomes
\begin{equation}
 \qquad \qquad \frac{\partial \psi}{\partial \tau}   =   H_{I} \, \psi
\label{highpolho2}
\end{equation}
wit $ H_{I}$  the harmonic oscillator Hamiltonian in coordinate space 
\begin{equation}
 H_{I} \equiv      \frac{1}{2} \Omeg \left( \lambda^{-2} \frac{\partial^2 }{\partial x^2}   -   \lambda^2 \, x^2 \right)
\label{highpolho3}
\end{equation}
$ H_{I}$ can be written in terms of  raising and lowering (ladder) operators  by using the diagonalizing $SL(2, \mathbb{R})$ matrix \eqref{diagonalM}
\begin{equation}
 H_{I} =     \frac{1}{2} \Omeg \, ( a^{\dagger} \, a + a \, a^{\dagger})
\end{equation}
with 
\begin{equation}
 a \equiv      \frac{1}{\sqrt{2}} (\lambda^{-1} \hat{p} - \lambda \, \hat{x}) \qquad  a^{\dagger} \equiv      \frac{1}{\sqrt{2}} (\lambda^{-1} \hat{p} + \lambda \, \hat{x}) 
\end{equation}
where
\begin{equation}
 \hat{p} = \frac{\partial }{\partial x} \qquad \quad  \hat{x} = x   
\end{equation}
Note that $a, a^{\dagger}$ are the regular ladder operators that are used for the construction of the harmonic oscillator Fock space in quantum mechanics.
The Heisenberg-Weyl commutation relations $[ \hat{p} , \hat{x}] = 1$  imply the commutator $[a,  a^{\dagger}] = 1$ since  $SL(2, \mathbb{R})$ is an automorphism of the  Heisenberg-Weyl  subgroup 
(equation\eqref{auto1}).
\vskip 1mm
\subsubsection{Polarized Functions in Coordinate Space}
\vskip 1mm
It is well know that the solution of equation \eqref{highpolho2}  can be represented as a series in Hermite functions $\varphi _{n}$ 
\begin{equation}
\psi(\tau, x) =   \sum_{n=0}^{\infty} e^{-\Omeg (n +\frac{1}{2}) \tau} \, \Alp_n \, \varphi_n (\lambda x)
\label{highpolho3}
\end{equation}
with $\Alp_n \in \mathbb{R}$ the  expansion coefficients.
The Hermite functions $\varphi _{n}$  have the following expression(\cite{abra})
\begin{equation}
\varphi _{n}(x)  = \frac{1}{\sqrt {2^{n}n!{\sqrt {\pi }}}} H_{n}(x) e^{-\frac{1}{2} x^{2}} 
\label{varphi}
\end{equation}
where $H_n$ are the Hermite polynomials.  $\varphi _{n}$  form an orthonormal basis in  $L_2(\mathbb{R})$ and fulfill the eigenvalue condition 
\begin{equation*}
\left(\frac{\partial^2 }{\partial x^2}   - \, x^2 \right) \, \varphi _{n }(x)= -(2 n +1 ) \varphi _{n}(x)  \nonumber
\end{equation*}
\vskip 1mm
The relationship between the (first) polarized functions \eqref{coherent2} and the polarized functions \eqref{highpolho3} is given by a modified Bargmann transform
\begin{align}
\varphi _{n} (\lambda \, x) &  =  \frac{1}{\sqrt {n!{\sqrt {\pi }}}} \times \nonumber \\
 &  \frac{1}{2 \pi i} \, \int _{-i \infty}^{i \infty } \, dp \,  \int _{- \infty}^{ \infty }  \, d x \,   A^n\, e^{-B^2/2 + \sqrt{2} \lambda \,x  B - \lambda^2 \, x^2/2} \, e^ {-A\, B} 
 \end{align}
where $A=A(p,q)$ and $B=B(p,q)$ are the variables defined in \eqref{diagonalFrac}. The integral over the momentum $p$ is an inverse double sided Laplace transform.
\vskip 1mm
\subsection{Pricing Kernel}
\vskip 1mm
The 
the orthogonality of the expansion \eqref{highpolho3} makes possible to express
the pricing kernel  $K_I(x,x^\prime, \tau)$, defined by
\begin{equation*}
\Psi(\tau, x)   = \int K_I(x,x^\prime, \tau) \Psi(0, x^\prime) d x
\end{equation*}
as a sum of products of Hermite polynomials in $x$ and $x^\prime$
\begin{align}
& K_I(x,x^\prime, \tau)  = \lambda \,  e^{-\frac{1}{2} \Omeg \tau} \, \sum_{=0}^{\infty}   \frac{ (e^{-   \Omeg\,\tau})^n} { 2^n n!}  \nonumber \\
& \qquad  \qquad  \,\times \,  H_n (\lambda x) \, H_n(\lambda  x^\prime )  \,  e^{ \frac{1}{2} \lambda^2  \, (x^2 + {x^\prime}^2)} 
\label{mehler1}
\end{align}
By applying the {\it{Mehler formula}} (\cite{mehler})
\begin{align*}
& \sum_{=0}^{\infty}   \frac{ \rho^n}{ 2^n n!} \, H_n(x) H_n(y)  \exp(\frac{1}{2} (x^2 + y^2)) =  \nonumber \\
& \quad \frac{1}{\sqrt{1-\rho^2}} \, \exp \left( \frac{4 \rho \, x \, y- (1+ \rho^2) (x^2 + y^2)} { 2 \, (1- \rho^2)} \right)  \nonumber
\end{align*}
 to the equation \eqref{mehler1},  with $\rho = \exp(- \Omeg \tau)$, one obtains the 
Mehler kernel, a  generalized bivariate Gaussian probability density
 \begin{equation}
K_I(x,x^\prime, \tau)  = \frac{\lambda}{\sqrt{ 2 \, \pi \, \sinh \Omeg \tau  }} \, \exp({  \frac{1}{2}  \,\lambda^2 \,  \bf{x} {\bf{A}} {\bf{x}}^ \intercal })
\label{mehlerK}
\end{equation}
where we have made the change $\tau =T -t$, $\tau \ge 0$, and  ${\bf{x}} \equiv (x, x^\prime)$. $\bf{A}$ is the $SL(2, \mathbb{R})$ matrix
\begin{equation}
\qquad \quad {\bf{A}} =  \begin{pmatrix*}[c]
    -\coth  \Omeg \tau &    \phantom{-}\csch  \Omeg \tau \\
     \phantom{-}\csch  \Omeg \tau  &  -\coth  \Omeg \tau 
\end{pmatrix*} 
\end{equation}
with $\coth$ and $\csch$ the hyperbolic cotangent and cosecant, respectively.  
\vskip 1mm
$\bf{A}$ is singular when $\Omeg \rightarrow 0$. However 
\begin{equation*}
\lim_{\Omeg \to 0} \Omeg \, \, {\bf{A}} =  \frac{1}{\tau} \, \begin{pmatrix*}[l]
  -1&       \phantom{-}1\\
     \phantom{-} 1&   -1
\end{pmatrix*}  
\end{equation*}
As expected from the group law \eqref{cocyclebs1z}, the Mehler kernel  maps into the heat kernel  (hence the Black-Scholes theory) when  $\Omeg \rightarrow 0$
\begin{equation}
\lim_{\Omeg \to 0} K_I(x,x^\prime, \tau)  = \frac{1}{\sqrt{2 \pi \Sig^2 \, \tau}} \, e^{ \frac{1}{2 \Sig^2 \tau}\,  (x - x^\prime)^2}
\end{equation}
\vskip 1mm
\subsubsection{Derivation using LCTs}
\vskip 1mm
The Mehler kernel \eqref{mehlerK} can also be obtained using the LCT associated  (\cite{Miller2}, \cite{wolf2} ) to the  harmonic oscillator generating matrix  \eqref{genmatrixHO}.
From \eqref{ltcdef2}
\begin{equation}
 W({\bf{M_H}},x,x^\prime)  = \frac{1}{\sqrt{2 \pi b}}\, \exp(  -( a {x^\prime}^2 - 2 x x^\prime + d\, x^2)/(2 \,b))
 \end{equation}
where $a,b,c,d$ refer to the elements of \eqref{genmatrixHO}. By direct substitution, we get
 \begin{equation}
 W({\bf{M_H}},x,x^\prime)  = K_I(x,x^\prime, \tau)  
\end{equation}
\vskip 1mm
A discussion of heat kernels and Mehler-type formulas based on group-invariant solutions can be found in 
reference \cite{mehler4}. Chapter 9 of \cite{wolf2} gives Baker-Campbell-Hausdorff relations between pseudo-differential  operators for the harmonic oscillator based on
composition of $SL(2, \mathbb{R})$ matrices.
\vskip 1mm
\subsection{Financial Interpretation}
\vskip 1mm
Using the Feynman-Kac formula,  we can write the solution of  \eqref{highpolho} as
\begin{equation}
\psi(x,t) = \mathbb{E} ( e^{ -  \frac{1}{2} \gamma  \int_t^{T} X(s) ^2 \, ds }\left.  \psi(X_T, T) \right|  X(t) = x)
\label{fkac2}
\end{equation}
with  $ \gamma \equiv \Omeg^2 / \Sig^2$
and where the expectation is taken with respect to a normal process with volatility $\sigma$. This equation describes a derivative with a payoff that is discounted quadratically
with the oscillator level $x$. A drift term can be easily added with a coboundary generated by $x$.
\vskip 1mm
A more interesting approach is to use the harmonic oscillator as a process describing stocks with non-normal returns with a correlation given by the Mehler Kernel  \eqref{mehlerK}.
The reference \cite{hos1} proposes a quantum harmonic oscillator as a model for the market force which draws a stock return
from short-run fluctuations to the long-run equilibrium.



%% file: repulsive_oscillator.tex
\vskip 1mm
Using the arguments in section \ref{OSGROUP} for the harmonic oscillator, because of the
$\mathbb{R}^{+}$ central extension in the Heisenberg-Weyl subgroup of $WSp(2,  \mathbb{R} )$,
the quantization group  for the repulsive oscillator $\tilde{H_R}$ is  generated by the elliptic  $SL(2, \mathbb{R})$ matrix
 \begin{equation}
 {\bf{M_R}} =  \begin{pmatrix*}[c]
    \phantom{-\lambda^{2}}  \cos \Omeg t &  \, \lambda^{-2}  \sin \Omeg t\\
    -\lambda^{2} \sin \Omeg t  &  \phantom{ \lambda^{-2}  } \cos\Omeg t
\end{pmatrix*} 
\label{genmatrixHOR}
\end{equation}
where $\Omeg, \Sig, t \in \mathbb{R} , \lambda \equiv \sqrt{\Omeg} /\Sig$
\vskip 1mm
 ${\bf{M_R}}$ is obtained from  the harmonic oscillator generating  matrix \eqref{genmatrixHO} by the mapping 
\begin{equation}
\qquad \Omeg \mapsto i \, \Omeg
\label{maprepulsive}
\end{equation}
For brevity, in this article we will not apply the full Group Quantization scheme to the repulsive oscillator.
Most results for the repulsive oscillator can be readily obtained from the harmonic oscillator results  and the correspondence \eqref{maprepulsive}.
\vskip 1mm
The repulsive oscillator group composition law is
\begin{align}
 t^{\prime \prime} & =t + t^\prime  \nonumber \\
p^{\prime \prime}& = p + p^\prime \,  \cos \Omeg t  +  \lambda^{2}   x^\prime \sin \Omeg t  \nonumber \\
 x^{\prime \prime} &  = x +   x^{\prime}   \, \cos \Omeg t  + \lambda^{-2}  \, p^\prime \,  \sin \Omeg t     \nonumber \\
 \zeta^{\prime \prime} & = \zeta^\prime \zeta \exp \left (  \epsilon_R(g^\prime, g) \right)
\label{cocyclebs1zr}
\end{align}
with the cocycle  $\epsilon_{R}$ 
\begin{align}
\epsilon_{R}(g,g^{\prime}) & = \frac{1}{2} ( p x^\prime - x p^\prime) \cos \Omeg t+ \\ \nonumber
 & \quad \frac{1}{2} (\lambda^{-2} p \, p^\prime - \lambda^{2} \, x \,  x^\prime)  \sin \Omeg t 
 \label{cocyclehs2r}
\end{align}
\vskip 1mm
One can easily verify that the high-polarization equation for the repulsive oscillator is (compare with  \eqref{highpolho})
\begin{equation}
 \frac{\partial \psi}{\partial t}  + \frac{1}{2} \Sig^2 \frac{\partial^2  \psi}{\partial x^2}   +  \gamma \, x^2 \, \psi = 0   \qquad  \quad \gamma \equiv \frac{\Omeg^2} {\Sig^2}
\label{highpolhor}
\end{equation}
whose separable solutions are time exponentials  and  Parabolic cylinder functions in $x$  (\cite{abra},  \cite{wolf2}, \cite{weber},  \cite{wolf10}).  Note that upon
the analytic continuation \eqref{maprepulsive} the Hermite functions \eqref{varphi} turn into Parabolic cylinder functions.
\vskip 1mm
The quantum repulsive oscillator \cite{barton} is a well known model in the literature.
A complete survey of the repulsive oscillator in the context of $WSp(2,  \mathbb{R} )$ can be found in \cite{wolf2}.  Chapter 9 of \cite{wolf2} 
gives special Baker-Campbell-Hausdorff relations between pseudo-differential  operators for the repulsive oscillator.
\vskip 1mm
\subsection{Financial Interpretation}
\vskip 1mm
Using the Feynman-Kac theorem, the solution of  \eqref{highpolhor} can be written as
\begin{equation}
\psi(x,t) = \mathbb{E} ( e^{   \frac{1}{2} \gamma  \int_t^{T} X(s)^2 \,  ds }\left.  \psi(X_T, T) \right|  X(t) = x)
\label{fkac3}
\end{equation}
where the expectation is taken with respect to a normal process with volatility $\sigma$. This equation describes a derivative with a payoff that grows quadratically
with the oscillator level $x$.
\vskip 1mm
Reference \cite{repul1} uses a repulsive anharmonic oscillator model to explain the distribution of financial
returns in a stock market when the market exhibits an upward trend. 

%% file: geometric.tex
\appendix

\section{Definitions}\label{DEF}
\vskip 1mm
\subsection*{Vector Fields}
\vskip 1mm
Let M be a N-dimensional manifold with local coordinates $x_i , i = 1,2 \ldots N$. A vector field $X$
is an application that associates a first order differential operator $X(x)$ to a point $x\in M$. 
$X(x)$ can be expressed in local coordinates as a linear combination of the base fields\
\begin{equation*}
e_i \equiv \frac{\partial}{\partial x_i} \qquad   i = 1,2 \ldots N
\label{fields1}
\end{equation*}
that is
\begin{equation*}
X(x) =  X_i(x) \frac{\partial}{\partial x_i}
\label{fields2}
\end{equation*}
with $X_i(x), i = 1,2 \ldots N$ differentiable functions on M. The space of the vector fields is called the
tangent space of M, T (M). For brevity we will write $X$  instead of $X(x)$

The integral curves of X are the solution to the set of ordinary differential equations
\begin{equation*}
\frac{d x_i}{d s} =  X_i(x) 
\label{fields3}
\end{equation*}
where s is an integration parameter. Note that the invariance condition $X f = 0$  implies that$f$ is constant along the integral curves of X.
\subsection*{Forms}
\vskip 1mm
A 1-form $\Gamma$  is an application that associates to every point $x \in M$ an element of the dual space of T (M).
\begin{equation*}
\Gamma: x \rightarrow \Gamma(x)  \qquad / \quad \Gamma(x) (X(x)) = f(x)
\label{forms1}
\end{equation*}
with $f(x)$  a differentiable function. 

As with vector fields, we write $\Gamma$ for $\Gamma(x)$. The space of 1-forms
is called the cotangent space of M and is denoted by $T^{*}(M)$.
A convenient representation for the basis of  $T^{*}(M)$ is
\begin{equation*}
u_i \equiv d x_i   \qquad   i = 1,2 \ldots N
\label{forms2}
\end{equation*}
and its action on the basis of T (M) is
\begin{equation*}
u_i (e_j) =\equiv d x_i (  \frac{\partial}{\partial x_j}) =  \delta_{i,j}     \qquad   i ,j = 1,2 \ldots N
\label{forms3}
\end{equation*}
2-forms, 3-forms , etc. are linear combinations of tensor products of 1-forms. A function f is considered a zero-form.

An important operation on forms is the differential d. The differential of a n-form is either a  (n+1)-form
or zero. For a function f, d is the ordinary differential
\begin{equation*}
d f(x) =  \frac{\partial f }{\partial x_i} d x_i
\label{forms4}
\end{equation*}
For a 1-form $\Gamma$
\begin{equation*}
d \Gamma = \frac{\partial \Gamma_i }{\partial x_j} d x_i \wedge d x_j
\label{forms5}
\end{equation*}
where the wedge operator $\wedge$ is the antisymmetric combination
\begin{equation*}
 d x_i \wedge d x_j = d x_i \otimes d x_j - d x_j \otimes d x_i 
\label{forms6}
\end{equation*}
One can define in an analogous way differentials of higher order forms. Note that the antisymmetry
of the wedge operator implies that $d^2 f = 0$ and $d^2 \Gamma = 0$. Also, i f a $n$-form acts on $n-k$ k vector fields one obtains a $k$-form. For instance the $1$-form $d f$ acting
on a field X gives a zero-form
\begin{equation*}
d f(X) = \frac{\partial f }{\partial x_i} d x_i ( X_j(x) \frac{\partial}{\partial x_j}) =  X_i(x) \frac{\partial f }{\partial x_i} =  X(f)
\label{forms7}
\end{equation*}
Analogously, one can check that $d \Gamma$ acting on X alone gives a 1-form
\begin{equation*}
d \Gamma(X,.) = \frac{\partial \Gamma_i}{\partial x_j}( X_i d x_j - X_j d x_i)
\label{forms8}
\end{equation*}
We use the inner product notation to denote the action of a n-for $\Omega$ on a vector field X
\begin{equation*}
i_X(\Omega) = \Omega(X)
\label{forms9}
\end{equation*}
\vskip 1mm
\subsection*{Lie Derivative}
\vskip 1mm
The Lie derivative evaluates the change of vector fields and forms
along the flow defined by another vector field.
\vskip 1mm
The Lie derivative of a function f with respect to a vector field X is
\begin{equation*}
L_X f  = X(f) = d f (X) 
\label{lie1}
\end{equation*}
For two vector fields X, Y , the Lie derivative of Y with respect to X, $L_X Y$ , is a vector field defined by
\begin{align*}
L_X Y  & =  - L_Y X \equiv  [X,Y] = \\
& \left(  X_i \frac{\partial Y_j}{\partial x_i}    -    Y_i \frac{\partial X_j}{\partial x_i}       \right)    \frac{\partial} {\partial x_j}
\label{lie2}
\end{align*}
The Lie derivative of a 1-form $\Gamma$ with respect to a vector field X,$L_X \Gamma$   is also a 1-form, and has the meaning of
the rate of change of $\Gamma$  along the integral lines (flow lines) of X. One finds, in local coordinates
\begin{equation*}
L_X \Gamma  = (L_X \Gamma)_i d x_i 
\label{lie3}
\end{equation*}
where
\begin{equation*}
 (L_X \Gamma)_i =     X_j \frac{\partial \Gamma_i}{\partial x_j}    +    \Gamma_j  \frac{\partial X_j}{\partial x_i}       
\end{equation*}
or, in a more concise notation
\begin{equation*}
L_X \Gamma  =  d (\Gamma(X)) + d\Gamma (X,.) \equiv i_X d\Gamma + d(i_X \Gamma)
\label{lie4}
\end{equation*}
\vskip 1mm
\subsection*{Lie Algebra}{\label{LieAlgebra}}
\vskip 1mm
A Lie algebra ${\mathfrak{G}}$  is a vector space over a field ${\mathbb{F}}$ equipped with  a bilinear map $[,]: ({\mathfrak{G}}, {\mathfrak{G}}) \rightarrow  {\mathfrak{G}}$
such that $[X,Y] = - [Y,X]$ and $[X,[Y,Z]] + [Y,[Z,X]]+ [Z,[X,Y]] = 0$ (Jacobi identity).This map is called  {\it{Lie bracket}}.
The Lie algebra for a Lie group G is defined using the tangent vectors at the identity $e$ as 
vector space of  and the Lie derivative  as bracket operator. One has
\begin{equation*}
 [X_{i}, X_{j}]  = c_{i,j}^{k}X_{k}^{L}  \qquad  i, j = 1,2 \ldots {\text{dim}} \,  G
\label{lieb0}
\end{equation*}
where the coefficients
$c_{i,j}^{k} \in  {\mathbb{F}},  i, j, k  = 1,2 \ldots {\text{dim}}\, G $ are called {\it{ structure constants}}. 

Since $X_g$ is a LIVF if and only if $X_g = {L_g^{T}}  X_{e}$ 
and an RIVF if and only if $X_g = {R_g^{T}}  X_{e}$, 
there is a biunivocal correspondence between the LIVFs (RIVFs)  and the set of the tangent vectors at the identity $e$ of $G$, with the following brackets
\begin{align}
 [X_{i}^{L}, X_{j}^{L}] & = c_{i,j}^{k}X_{k}^{L} \qquad [X_{i}^{R}, X_{j}^{R}] = - c_{i,j}^{k}X_{k}^{R} \nonumber \\
[X_{i}^{R}, X_{j}^{L}] & = 0 \qquad \forall \, i, j = 1,2 \ldots {\text{dim}} \,  G
\label{lieb1}
\end{align}
\vskip 1mm

%% file: black_scholes_prices_in_laplace_space.tex
\section{Instrument Prices  in Momentum Space} \label{BRAKET}
\vskip 1mm
In this section we present examples of  derivatives pricing in the less familiar momentum (Laplace) space. 
\vskip 1mm
The price of a financial instrument $ \Psi ( x,  t)$  maturing at $T$ is found by performing an inverse Laplace transform 
\begin{equation}
 \Psi ( x,  t)  =  \frac{1 }{2 \pi i} \, \int _{c-i \infty}^{c+ \infty } \, d p \,   e^{x p}  K(p,\tau) \Psi(p,T)  
\label{price1}
\end{equation}
where $\tau \equiv T-t, \, t \leq T$ and $\Psi(p,T)  =    \Psi(p)  \exp(E(p) T)$  is the payoff at maturity in the Laplace space
\begin{equation}
\Psi ( x,  T)  =  \frac{1 }{2 \pi i} \, \int _{c-i \infty}^{c+ \infty } \, d p \,   e^{x p}  \Psi(p,T)  
\label{price2}
\end{equation}
\vskip 1mm
\subsection{Instrument Prices using the Mellin Transform}
\vskip 1mm
The inverse Mellin transform{\footnote{${\mathcal {M}}^{-1}$ exists only for
complex values of $y$  so that $c \equiv \Re(y) > 0$ is within certain (possibly multiple)  {\it{convergence strips}}. Each strip leads to different results for ${\mathcal {M}}^{-1}(f)$
(\cite{Mellin2},  \cite{Mellin2b}, \cite{Mellin4},  \cite{Mellin5}).}}  gives the  instrument's price in terms of the stock value instead than the log-stock. The pricing equations \eqref{price1} and \eqref{price2} read now
\begin{align}
 V ( S,  t)  & =  \frac{1 }{2 \pi i} \, \int _{c-i \infty}^{c+ \infty } \, d p \,  \exp( - E(p) \tau) V(p,T)  \,   S^{p}   \nonumber \\
V ( S,  T)  & =     \frac{1 }{2 \pi i} \, \int _{c-i \infty}^{c+ \infty } \, d p \,   S^{p}  V(p,T)  
\label{mellinprice}
\end{align}
with
\begin{equation*}
 E_{r}(p) \equiv \frac{1}{2 } \sigma^2 \,p^2 + \mu p -r \qquad \mu = 1- \frac{1}{2 } \sigma^2
 \end{equation*}
 The final condition in Mellin space is given by
\begin{equation}
C(p,0)  =   \int _{0}^{+ \infty } V ( S,  T) \, S^{-p-1} dS  
\label{mellinfc}
\end{equation}
\vskip 1mm
\subsection{Call Option Price}
\vskip 1mm
We price a call option  with strike $X$  and maturity $T$ using the Mellin transform.
The payoff  is, from \eqref{price2}
\begin{equation*}
 C(p,0)  =   \int _{0}^{+ \infty } (S-X)^{+} S^{-p-1} dS  =   \frac{X^{1-p}}{p (p-1)}
\end{equation*}
with $ \Re(p) > 1$. The expression of the call option price in the Mellin space is, using \eqref{mellinprice}
\begin{equation*}
 C(p,\tau)  =  e^{- r \tau} \,  \frac{X^{1-p}}{p (p-1)} \,    \exp ( - ( \frac{1}{2 } \sigma^2 \,p^2 + \mu p)   \tau)
\end{equation*}
where $\tau \equiv T-t, \, t \leq T$. 
\vskip 1mm
The Black-Scholes call option formula in the price (coordinate)  representation is obtained by the inverse Mellin transform
\begin{equation*}
 V(S, t)   =   e^{- r \tau}   \frac{1 }{2 \pi i} \, \int _{c-i \infty}^{c+ \infty } \,    C(p,\tau)  \,  S^p \, d p
\label{call00}
\end{equation*}
where $c \in (1, +\infty)$  and $ \Re(p) > 1$. 
\vskip1mm
Details of the calculation of this integral can be found in reference \cite{Mellin0}.

%% file: black_scholes_lagrangian.tex
\
\section{Lagrangian Formulation}\label{CLASSA}
In the Group Quantization formalism,
$ \Theta \rvert_{\mathcal{C}}$,  the projection onto the base manifold of the vertical form $\Theta$ along the integral trajectories of the
 characteristic module $C_\Theta$ provides {\footnote{
If the quantization group $\tilde{G}$ is finite, the quotient space  $P = \tilde{G}/\mathcal{C}_\Theta$ and the quotient connection form  $\Theta_P = \Theta /\mathcal{C}_\Theta$, 
where $\mathcal{C}_\Theta$ is the characteristic module of  $\Theta$,
define a fiber bundle $(P,U, \pi)$   where the curvature form $\Omega = d \Theta_P$ is a symplectic form over $P/U$. Taking the quotient by the integral flows of $\mathcal{C}_\Theta$ allows the definition of 
local coordinates in $P/U$,   $(p_i,q_i), i=1,2 \ldots \frac{1}{2} dim(P/U)$  representing the canonical momenta and canonical coordinates.
of the vertical form $\Theta$ along the integral flows (trajectories) of the
 characteristic module $C_\Theta$.}}
the classical action ${\mathcal{S}}$  (\cite{aldaya0} ) 
\vskip 1mm
\begin{equation*}
{\mathcal{S}} =    \int  \Theta \rvert_{\mathcal{C}} =   \int  {\mathcal{L}} (x, \dot{x}) d t 
\label{act1}
\end{equation*}
The classical Lagrangian ${\mathcal{L}} (x, \dot{x})$ is obtained by the projection onto the base manifold. Note that $\Theta$ and $ \Theta + d f$  generate equivalent Lagrangians that differ in a total derivative and do not change the action ${\mathcal{S}}$
\vskip 1mm
The vertical (connection) form  $d \Theta$  restricted to the quotient space  ${\tilde{G}}/U$
 is analogous to the {\it{Poincar{\'e}-Cartan}} form of classical mechanics $\Theta_{PC}$ ,which can
 be written in local coordinates ${\bf{q}}, {\bf{p}}$ as
\begin{equation}
\Theta_{PC} = \sum_i  p_i d q_i - H({\bf{p}}, {\bf{q}}) dt
\label{mod1bb4pb}
\end{equation}
where $H({\bf{p}}, {\bf{q}})$ is the Hamiltonian function, and ${\bf{q}}, {\bf{p}}$ are conjugate pairs of coordinates and momenta. In classical mechanics,
the equations of motion are just the flows associated to the Hamiltonian field 
\begin{equation*}
X_H \equiv  \frac{\partial}{\partial t} +  \sum_i  \frac{\partial H}{\partial p_i} \frac{\partial}{\partial q_i} - \frac{\partial H}{\partial q_i} \frac{\partial}{\partial p_i}
\label{mod1bb66pb}
\end{equation*}
 so along Hamiltonian trajectories 
\begin{equation*}
{\dot{q_i}} \equiv \frac{d q_i}{ d t } =  \frac{\partial H}{\partial p_i} \qquad {\dot{p_i}}\equiv  \frac{d p_i}{d t } = - \frac{\partial H}{\partial q_i} 
\label{mod1bb6ppp}
\end{equation*}
Note that $\Theta_{PC}$ has the structure of a Legendre transformation 
if one identifies $p_i \equiv \partial {\mathcal{L}} / \partial {\dot{q_i}}$, where ${\mathcal{L}}$ is the Lagrangian function. Along the Hamiltonian flows 
\begin{equation*}
\Theta_{PC}  \rvert_{H}=  \mathcal{L}({\dot{{\bf{q}}}}, {\bf{q}}) dt 
\label{mod1bb4ppk}
\end{equation*}
The differential of  $ \Theta \rvert_{PC}$  along the Hamiltonian trajectories  is a symplectic form.
\begin{equation*}
 d \Theta_{PC}   \rvert_{H} =  \sum_i  d p_i \wedge d q_i
\label{mod1bb4}
\end{equation*}
The Lagrangian function is used for the application of  {\it{path integral}} methods in computational finance  ( \cite{baaquie1},  \cite{baaquie4}, \cite{baaquie3}, \cite{linetsky}).
\vskip 1mm
\subsection{Black-Scholes Lagrangian}\label{LAGBS}
We find the Black Scholes Lagrangian by restricting the  Black-Scholes connection $\Theta$ \eqref{theta2bs}  to the characteristic module trajectories on the base manifold.
\vskip 1mm
$\mathcal{C}_\Theta$ is  generated by the time translations ${X_{t}^L}$ whose integral flows  on the base manifold are (see  \eqref{mod2bbbs})
\begin{align}
& \frac{dt} { ds}  = 1   \quad \frac{dp} { ds}  = 0  \quad    \frac{dx} { ds} = \sigma^2 p  + \mu        \nonumber \\
& \rightarrow \quad t  = s         \quad  p = p_0   \quad         x =  x_0 + (\sigma^2  p_0 + \mu )\, t  
\label{mod3bs}   
\end{align}
where $x_0$, $p_0$ , $x_0$, $\zeta_0$    are  integration constants and $s$ is the integration parameter. 
\vskip 1mm
We first add the total differential $d (p x) $ to $\Theta$ so the equations adopt the Poincar{\'e}-Cartan expression \eqref{mod1bb4pb}
\begin{equation}
\Theta  \rightarrow  d(x p) + \Theta  =  p \, dx -  E(p) \, dt   
\end{equation}
then
\begin{align*}
 \Theta \rvert_{\mathcal{C}} & =   p  (\sigma^2 p  + \mu) d t  - \left(  \frac{1}{2 } \sigma^2 p^2 + \mu p  -r \right) d t  =  \nonumber \\
 &   \frac{1}{2 \sigma^2} ( \dot{x} - \mu)^2   dt  - r dt
\label{theta4bs}
\end{align*}
where we have used equation \eqref{mod3bs} 
\begin{equation*}
\frac{dx} { dt} \equiv  \dot{x} = \sigma^2 p  + \mu  \rightarrow    p = \frac{ \dot{x} - \mu}{\sigma^2} 
\end{equation*}
with the result
\begin{equation}
 \mathcal{L}  (x, \dot{x}) =    \frac{1}{2 \sigma^2} ( \dot{x} -  \mu)^2 - r 
\label{theta5bsa}
\end{equation}
The solution $x(\tau)$ of the Lagrangian \eqref{theta5bsa} coincides with the expected value of the coordinate (log-price) using the
$K_{BS}$  kernel in \eqref{kernellag1}) 
\begin{equation}
\langle  \hat{x}  \rangle =   \, \frac{1}{\sqrt{2 \pi \sigma^2 \tau}} \, \int _{-\infty}^{ \infty } \, d x  \, x   \, \ e^{ -\frac{1}{2 \sigma^2 \tau } \, (x^\prime-x -\mu \tau)^2} = x^\prime + \mu \tau
 \label{kernelBS1}
\end{equation}
\vskip 1mm
\subsection{Euclidean Oscillator Lagrangian}\label{LAGHOH}
\vskip 1mm
Following the same steps than in section \ref{LAGBS}, and using the expressions \eqref{theta2bs} and \eqref{energyHO}, we find that the $\mathcal{C}_\Theta$  trajectories
\begin{equation}
 \frac{dt} { ds}  = 1   \quad \frac{dp} { ds}  =  \Omeg \, \lambda^2 x   \quad    \frac{dx} { ds} = \Omeg \, \lambda^{-2} p     
\end{equation}
generates the following Lagrangian for the Euclidean harmonic oscillator
\begin{equation}
 \mathcal{L}_{H}  (x, \dot{x}) =    \frac{1}{2\sigma^2}\,  \left(  {\dot{x}}^2 +    \frac{1}{2} \Omeg^2 \, x^2\right)
\end{equation}
\vskip 1mm
As expected, the mapping $\Omeg \rightarrow i \Omeg$ provides the Lagrangian for the Euclidean repulsive oscillator
\begin{equation}
 \mathcal{L}_{R}  (x, \dot{x}) =     \frac{1}{2\sigma^2}\, \left( {\dot{x}}^2 -  \frac{1}{2}   \Omeg^2 \, x^2 \right)
\end{equation}

%% file: acknowledgements.tex
The author is grateful to Peter Carr for very many valuable insights and  support, and to Gregory Pelts for sharing his knowledge on the geometric structure of finance.
\vskip 3mm

\section*{Disclaimer}
\vskip 3mm

The views expressed herein are those of the author and do not reflect the views of my current employer, Wells Fargo Securities, or affiliate entities.

%% file: main_4.bbl
\begin{thebibliography}{9}

\bibitem{aldaya0}
V. Aldaya and J. A. de Azcarraga, 
\textit{Quantization as a consequence of the symmetry group: An approach to geometric quantization},
J. Math. Phys. {\bf{23}},
1982.


\bibitem{wolf11}
V. Aldaya and J. A. de Az\'carraga and K.B. Wolf
\textit{Quantization, symmetry, and natural polarization},
\textit{J. Math. Phys. 25 (3), March 1984},


\bibitem{garcia0}
V. Aldaya, J. A. de Azc\'arraga and S. Garc\'ia,
\textit{Group manifold approach to field quantisation},
Journal of Physics A: Mathematical and General, {\bf{21}}, 23, 
1988.
 
 

\bibitem{garcia1}
Santiago Garc\'ia,
\textit{Hidden invariance of the free classical particle},
\textit{https://arxiv.org/abs/hep-th/9306040}
American Journal of Physics, {\bf{62}}, 6,
1994. 



\bibitem{bisquert}
V.Aldaya, J.Bisquert, J.Guerrero, J.Navarro-Salas
\textit{Group-theoretical construction of the quantum relativistic harmonic oscillator},
Reports on Mathematical Physics
Volume 37, Issue 3,
June 1996, Pages 387-418,
1997.

\bibitem{aldaya22}
Aldaya V., Guerrero J., Marmo G. 
 \textit{Quantization on a Lie Group: Higher-Order Polarizations},
 In: Gruber B., Ramek M. (eds) Symmetries in Science X. Springer, Boston, MA,
\textit{arXiv:physics/9710002v1},
1998.

\bibitem{calixto}
 M. Calixto,  V.Aldaya; and M. Navarro,
\textit{Quantum Field Theory in Curved Space from a Second Quantization of a Group},
\textit{Int.J.Mod.Phys. {\bf{ A15}} },
\textit{https://arxiv.org/abs/hep-th/9701180},
 Mar 2000.

\bibitem{aldaya6}
 V. Aldaya,  J. Guerrero,
\textit{Lie Group Representations and Quantization},
Reports on Mathematical Physics, {\bf{47}}, 
2001.


\bibitem{Weinorman}
 J. Wei and E. Norman,
\textit{Lie algebraic solution of linear differential equations},
\textit{J. Math. Phys., 4(4):575–581,},
1963.


\bibitem{Weinorman2}
Szymon Charzynski and Marek Kus,
\textit{Wei-Norman equations for a unitary evolution},
\textit{Journal of Physics A: Mathematical and Theoretical, Volume 46, Number 26},
2013.
  
\bibitem{lo}
C. F. Lo,
\textit{Lie-Algebraic Approach for Pricing Zero-Coupon Bonds in Single-Factor Interest Rate Models},
Journal of Applied Mathematics,Volume 2013, Article ID 276238, 
2013.

\bibitem{Miller}
E.G. Kalnins, W. Miller, Jr., G.S. Pogosyan
\textit{Complete sets of invariants for dynamical systems that admit a separation of variables},
Journal of Mathematical Physics 43(7), pp.--3592-3609,
July 2002.

\bibitem{Miller2}
W. Miller, Jr.,
\textit{The Scrodinger and Heat Equations},
Encyclopedia of Mathematics and its Applications,Vol. 4,
Cambridge University Press, 
2010.


\bibitem{mehler4}
F. Gungor,
\textit{Equivalence and Symmetries for Linear Parabolic Equations and Applications Revisited},
\textit{arXiv:1501.01481 [math-ph]},
2017.

\bibitem{linetsky}
Vadim Linetsky,
\textit{The Path Integral Approach to Financial Modeling and Options Pricing},
Computational Economics,
{\bf{11}},1, 
1997.

\bibitem{abra}
Abramowitz, Milton, and Irene A. Stegun. 
\textit{Handbook of Mathematical Functions with Formulas, Graphs, and Mathematical Tables},
Dover Publications,
Ninth printing, 1970.


\bibitem{kozlov}
Roman Kozlov
\textit{On Lie Group Classification of a Scalar Stochastic Differential Equation},
 Journal of Nonlinear Mathematical Physics, {\bf{18}},1,
 2010.

\bibitem{pazy}
A. Pazy, 
\textit{Semigroups of Linear Operator and Applications to Partial Differential Equations},
\textit{Applied Mathematical Sciences, Springer-Verlag, New York},
1983.
group

\bibitem{Baker}
F. Soto-Eguibar and H. M. Moya-Cessa,
\textit{Solution of the Schrodinger Equation for a Linear Potential using the Extended Baker-Campbell-Hausdorff Formula},
Appl. Math. Inf. Sci. 9, No. 1, 175-181,


\bibitem{conformal1}
Juan M Romero, Elio Mart{\'i}nez Miranda and Ulises Lavana,
\textit{Conformal symmetry in quantum finance},
 J. Phys.  Conf. Ser. {\bf{512}},
 2014.
 
 
\bibitem{conformal4}
Juan M Romero, Elio Martínez Miranda and Ulises Lavana,
\textit{Schrodinger group and quantum finance}
\textit{https://arxiv.org/abs/1304.4995}


\bibitem{conformal3}	
U. Niederer,
\textit{The maximal kinematical invariance group of the free Schrodinger equation},
 Helv. Phys. Acta, {\bf{45}},
1972.

\bibitem{conformal2}
C.R. Hagen,
\textit{Scale and conformai transformations in galilean-covariant field theory},
 Phys. Rev. {\bf{5}},2,
 1972.


\bibitem{baaquie1}
Belal Baaquie,
\textit{Quantum Finance: Path Integrals and Hamiltonians for Options and Interest Rates},
Cambridge University Press,
2004.

 
\bibitem{baaquie2}
Belal Baaquie,
\textit{Interest Rates and Coupon Bonds in Quantum Finance},
Cambridge University Press,
2009.

 

\bibitem{baaquie4}
Belal Baaquie,
\textit{Financial modeling and quantum mathematics},
Computers and Mathematics with Applications 65 (2013) 1665–1673,


\bibitem{baaquie3}
Belal Baaquie,Claudio Coriano, Marakani Srikant,
\textit{Quantum Mechanics, Path Integrals and Option Pricing: Reducing the Complexity of Finance},
\textit{arXiv:cond-mat/0208191 [cond-mat.soft]},
2002.

\bibitem{lie1}
R. K. Gazizov, N. H. Ibragimov,
\textit{Lie Symmetry Analysis of Differential Equations in Finance},
Nonlinear Dynamics,{\bf{17}},4,
1998

\bibitem{zazen}
F. Casas,  A. Murua M. Nadini,
\textit{Efficient computation of the Zassenhaus formula},
Computer Physics Communications,
Volume 183, Issue 11, pp 2386--2391,
November 2012


\bibitem{wolf0}
K.B. Wolf,
\textit{On Time-Dependent Quadratic Quantum Hamiltonians},
\textit{SIAM Journal on Applied Mathematics, Vol. 40, No. 3, Jun., 1981, pp. 419-431}


\bibitem{wolf10}
C. Munoz, J. Rueda-Paz and K.B. Wolf,
\textit{Discrete repulsive oscillator wavefunctions},
\textit{J. Phys. A: Math. Theor. 42 (2009) 485210 (12pp)}


\bibitem{wolf2}
K.B. Wolf,
\textit{Integral Transforms in Science and Engineering},
Plenum,
1979


\bibitem{repul1}
N. Jaroonchokanan and S. Suwanna,
\textit{Inverted anhamonic oscillator model for distribution of financial returns},
\textit{ J. Phys.: Conf. Ser. 1144 012101},
2018

\bibitem{barton}
G. Barton,
\textit{Quantum mechanics of the inverted oscillator potential},
\textit{Ann. Phys. (NY) 166 322–63},
 1986 
 
\bibitem{weber}
A. Wunsche,
\textit{Associated Hermite Polynomials Related to Parabolic Cylinder Functions},
\textit{Advances in Pure Mathematics, 9, 15-42},
2019

\bibitem{lie2}
A Paliathanasis, RM Morris and PGL Leach,
\textit{Lie symmetries of 1+2 nonautonomous evolution equations in Financial Mathematics},
\textit{https://arxiv.org/pdf/1605.01071.pdf},
2016.


\bibitem{freefall}
Nicholas Wheeler, 
\textit{Classical/Quantum Dynamics in a Uniform Gravitational Field: A. Unobstructed Free Fall},
Preprint, Reed College Physics Department
August 2002

\bibitem{goldstein}
Goldstein, Robert S. and Keirstead, William P.,
\textit{On the Term Structure of Interest Rates in the Presence of Reflecting and Absorbing Boundaries },
 Available at SSRN: https://ssrn.com/abstract=19840 or http://dx.doi.org/10.2139/ssrn.19840,
June 1997

\bibitem{mehler}
 Mehler, F. G. ,
 \textit{Ueber die Entwicklung einer Function von beliebig vielen Variabeln nach Laplaceschen Functionen höherer Ordnung},
 Journal für die Reine und Angewandte Mathematik (in German) (66),
1866

\bibitem{bargmann1}
V. Bargmann, 
Ann. Math. {\bf{59}}, 1,
1954.


\bibitem{bargmann2}
V. Bargmann, 
Commun. Pure Appl. Math. {\bf{14}}, 187 (1961); {\bf{20}}, 1, (1967).


\bibitem{cartan}
\'Elie Cartan, 
\textit{Les vari\'et\'es \`a connexion affine et la th\'eorie de la relativit\'e g\'en\'eralis\'ee, I et II},
Ann. Ec. Norm. {\bf{40}}, 1923, and{ \bf{41}}, 1924.



\bibitem{geom1}
J. M. Souriau, 
\textit{Structure des systemes dynamiques}
 Dunod, Paris, 1970.
 
 \bibitem{geom2}
 B. Kostant,
 \textit{Quantization and Unitary Representations},
 Lecture Notes in Mathematics {\bf{170}}, 
 Springer-Verlag, New York, 
1970.

\bibitem{geom3}
D. J.Simms and N. M. J. Woodhouse, 
\textit{Lecture Notes in Geometric Quantization (Springer-Verlag}, 
Berlin, 1976.


\bibitem{Labord}
Pierre-Henry Labord\`ere,
\textit{Analysis, Geometry, and Modeling in Finance: Advanced Methods in Option Pricing},
\textit{Chapman and Hall/CRC Financial Mathematics Series},
2008.
 
\bibitem{jana}
T.K. Jana, P. Roy,
\textit{Pseudo Hermitian formulation of Black-Scholes equation},
\textit{https://arxiv.org/abs/1112.3217},
2011
 
 
 
\bibitem{boson}
N. Bebiano, J. da Providencia and J.P. da Providencia,
\textit{Non-Hermitian quantum mechanics of bosonic operators},
\textit{https://arxiv.org/abs/1705.11153},
2017

\bibitem{barut}
Asim O. Barut, Ryszard Raczka,
\textit{Theory of Group Representations and Applications},
\textit{Ars Polona},
1980


\bibitem{masao}
Masao Nagasawa,
\textit{Scrodinger Equations and Diffusion Theory},
\textit{Birkhauser-Verlag},
1993


\bibitem{paco1}
S. Garc\'ia and F. Garc\'ia, 
\textit{Imaginary Mass, Black-Scholes Variance, and Group Quantization (March 14, 2018)},
\textit{ Available at SSRN: https://ssrn.com/abstract=3140485 or http://dx.doi.org/10.2139/ssrn.3140485}


\bibitem{ozemir}
C. Ozemir and  F. Gungor,
\textit{Symmetry classification of variable coefficient cubic-quintic nonlinear Schrodinger equations},
\textit{arXiv:1201.4033v1},
2012.

\bibitem{lie1}
R. K. Gazizov, N. H. Ibragimov,
\textit{Lie Symmetry Analysis of Differential Equations in Finance},
\textit{Nonlinear Dynamics,{\bf{17}},4},
1998.

\bibitem{blasi}
A. Blasi, G. Scolarici‡and L. Solombrino,
\textit{Pseudo-Hermitian Hamiltonians, indefinite inner product spaces and their symmetries},
\textit{arXiv:quant-ph/0310106v2},
2004.

\bibitem{Mellin0}
Radha Panini, Ram Prasad Srivastav,
\textit{Option pricing with Mellin transnforms},
\textit{Mathematical and Computer Modelling,40,43-56},
 2004.

\bibitem{Carr0}
P. Carr, R. Jarrow, R. Myneni, 
\textit{Alternative characterizations of American put options}, 
\textit{Mathematical Finance 2, 87–105}.
1992


\bibitem{Mellin2}
J. Bertrand, P. Bertrand and  J. Ovarlez,
\textit{The Transforms and Applications Handbook: Second Edition.},
\textit{The Mellin Transform,Ed. Alexander D. Poularikas,Boca Raton: CRC Press LLC },
2000.

\bibitem{Mellin2b}
Paul L. Butzer
\textit{A Direct Approach to the Mellin Transform},
\textit{Journal of Fourier Analysis and Applications},
July 1997.

\bibitem{Mellin4}
R. Frontczak,
\textit{Pricing Options in Jump Diffusion Models Using Mellin Transforms.},
\textit{Journal of Mathematical Finance,3,366-373},
2013.

\bibitem{Mellin5}
P. Buchen,
\textit{An Introduction to Exotic Option Pricing},
\textit{CHAPMAN \& HALL/CRC, Financial Mathematics Series},
2012.

\bibitem{hos1}
K. Ahn, M.Y. Choi, B. Dai, S. Sohn and B. Yang,
\textit{Modeling stock return distributions with a quantum  harmonic oscillator},
\textit{EPL 120 38003},
\textit{https://doi.org/10.1209/0295-5075/120/38003},
2018

\end{thebibliography}
